\documentclass[12pt]{article}
\pdfoutput=1

\usepackage{putex}
\usepackage{amsmath,amssymb,amsfonts}
\usepackage{psfrag}
\usepackage{footmisc}
\usepackage{url}
\usepackage{mathtools}
\usepackage{color}

\usepackage{tikz}
    \usepackage{amssymb,amsfonts,amsmath}
    \usepackage{tkz-euclide}
        \usetikzlibrary{arrows,calc,patterns}
\usepackage{pgfplots}

\definecolor{darkblue}{rgb}{0.1,0.1,.7}
\definecolor{purple}{rgb}{0.6,0,0.6}
\definecolor{orange}{rgb}{0.9,0.6,0}
\usepackage[colorlinks, linkcolor=darkblue, citecolor=darkblue, urlcolor=darkblue, linktocpage]{hyperref} 
\usepackage[square, comma, compress,numbers]{natbib}
\usepackage[]{amsmath}
\usepackage[]{graphicx}
\usepackage[]{latexsym}
\usepackage[utf8]{inputenc}
\usepackage{geometry}
\usepackage{amscd}
\usepackage[all,cmtip]{xy}
\usepackage{mathrsfs}
\usepackage[margin=10pt,font=small,labelfont=bf]{caption}
\usepackage{dsdshorthand}
\usepackage{changepage}
\usepackage{setspace}
\usepackage{tikz}
\usepackage{subcaption}

\usepackage[titles]{tocloft}
\usetikzlibrary{arrows,positioning}

\def\ma{{\mathbf \alpha}}

\def\SL2{\widetilde{SL}(2,\mathbb R)}

\newcommand\mR{\mathbb{R}}
\newcommand\mZ{\mathbb{Z}}

\newcommand\lpl{\ell_\text{Pl}}

\numberwithin{equation}{section}

\newcommand{\grp}[1]{\mathrm{#1}}

\newcommand{\bdy}{\text{bdy}}
\newcommand{\extr}{\text{ext}}
\newcommand{\nextr}{\text{near-ext}}

\newcommand {\bes} {\begin {equation*}}
\newcommand {\ees} {\end {equation*}}
\newcommand {\beq} {\begin {equation}}
\newcommand {\eeq} {\end {equation}}
\newcommand {\bea} {\begin {eqnarray}}
\newcommand {\ea} {\end {eqnarray}}

\newcommand{\Sch}{\text{Sch}}

\numberwithin{equation}{section}

\def\<{\langle}
\def\>{\rangle}

\tikzset{
    >=stealth',
    punkt/.style={
           rectangle,
           rounded corners,
           draw=black, very thick,
           text width=15em,
           minimum height=2em,
           text centered},
    pil/.style={
           ->,
           thick,
           shorten <=2pt,
           shorten >=2pt,}
}

 \def\ie{\begin{equation}\begin{aligned}}
\def\fe{\end{aligned}\end{equation}}

\begin{document}

\begin{flushright}
\hfill{\tt PUPT-2615}
\end{flushright}

\institution{PU}{${}^1$ Joseph Henry Laboratories, Princeton University, Princeton, NJ 08544, USA}
\institution{UCSB}{${}^2$ Physics Department, University of California, Santa Barbara, CA 93106, USA}

\title{
The statistical mechanics of near-extremal black holes 
}

\authors{Luca V. Iliesiu${}^1$ and Gustavo J. Turiaci${}^2$}

\abstract{
An important open question in black hole thermodynamics is about the existence of a ``mass gap'' between an extremal black hole and the lightest near-extremal state within a sector of fixed charge. 
In this paper, we reliably compute the partition function of Reissner-Nordstr\"{o}m near-extremal black holes at temperature scales comparable to the conjectured gap. We find that the density of states at fixed charge does not exhibit a gap; rather, at the expected gap energy scale, we see a continuum of states. We compute the partition function in the canonical and grand canonical ensembles, keeping track of all the fields appearing through a dimensional reduction on $S^2$ in the near-horizon region. Our calculation shows that the relevant degrees of freedom at low temperatures are those of $2d$ Jackiw-Teitelboim gravity coupled to the electromagnetic $U(1)$ gauge field and to an $SO(3)$ gauge field generated by the dimensional reduction.    
}
\date{}

\maketitle
\tableofcontents

\newpage
\section{Introduction and outline}
\label{sec:intro}

Extremal and near-extremal black holes have long offered a simplified set-up to resolve open questions in black hole physics, ranging from analytic studies of mergers to microstate counting. The simplicity of near-extremal black holes comes from the universality of their near-horizon geometry: there is an $AdS_2$ throat with an internal space that varies slowly as the horizon is approached (see, for example, \cite{Kunduri:2007vf}).  

While the near-horizon geometry exhibits great simplicity, the thermodynamics of extremal and near-extremal black holes brings up several important open questions.  At extremality, black holes have zero temperature, mass $M_0$, and area $A_0$.  Performing a semiclassical analysis when raising the mass slightly above extremality, one finds that the energy growth of near-extremal black holes scales with temperature as $\delta E = E-M_0 = T^2/M_{\text{gap}}$. Naively, one might conclude that when the temperature, $T < M_{\text{gap}}$, the black hole does not have sufficient mass to radiate even a single Hawking quanta of average energy.  Consequently, $M_{\text{gap}}$ is considered the energy scale above extremality at which the semiclassical analysis of Hawking must breakdown \cite{Preskill:1991tb, Maldacena:1998uz, Page:2000dk}.\footnote{Even at temperatures $T \sim O(M_{\text{gap}})$ there is a  breakdown of thermodynamics since a single  Hawking quanta with average energy could drastically change the temperature of the black hole. } A possible way to avoid the failings of the semiclassical analysis is to interpret $M_{\text{gap}}$ as a literal ``mass gap'' between the extremal black hole and the lightest near-extremal state in the spectrum of black hole masses. Such a conjecture is, in part, supported by microscopic constructions \cite{Callan:1996dv, Maldacena:1996ds, Maldacena:1997ih} which suggest that, in the case of black holes with sufficient amounts of supersymmetry, $M_{\text{gap}}$ could indeed be literally interpreted as a gap in the spectrum of masses.\footnote{In \cite{Maldacena:1997ih}, it is assumed that the lightest near-extremal state has non-zero spin, in contrast to the extremal Reissner-Nordstr\"{o}m. However, in section \ref{sec:part-function-large-BHs-AdS4}, we show that in fact the lightest near-extremal state  has zero spin. \cite{Callan:1996dv, Maldacena:1996ds} focus on string constructions for near-extremal black holes in supergravity. Since our analysis depends on the massless matter content in the near-horizon region, we cannot compare our results with the gaped results of \cite{Maldacena:1996ds}. Nevertheless, an analysis of the $2d$ effective theory in the near-horizon region for near-extremal black holes in supergravity is currently underway \cite{HITZ}.     } Nevertheless, it is unclear if such results are an artifact of supersymmetry or whether such a gap truly exists for the most widely-studied non-supersymmetric examples: in Reissner-Nordstr\"{o}m (RN) or Kerr-Newman (KN) black holes.

The mass-gap puzzle is related to another critical question of understanding the large zero-temperature entropy of extremal black holes. If a gap exists, and the semiclassical analysis is correct even at low temperatures, extremal black holes would exhibit a huge degeneracy proportional to the macroscopic horizon area measured in Planck units. In the absence of supersymmetry, it is unclear how such a degeneracy could exist without being protected by some other symmetry. Alternatively, if one takes the semiclassical analysis seriously only at temperatures $T\gg M_{\text{gap}}$, then it is possible that the entropy obtained by this analysis would not count the degeneracy of the ground-state; rather, it could count the total number of states with energy below $E -M_0\lesssim M_{\text{gap}}$~\cite{Page:2000dk}. We find this solution unsatisfactory since, from the point of view of the Gibbons-Hawking prescription, we should be able to compute the Euclidean path integral at lower temperatures.

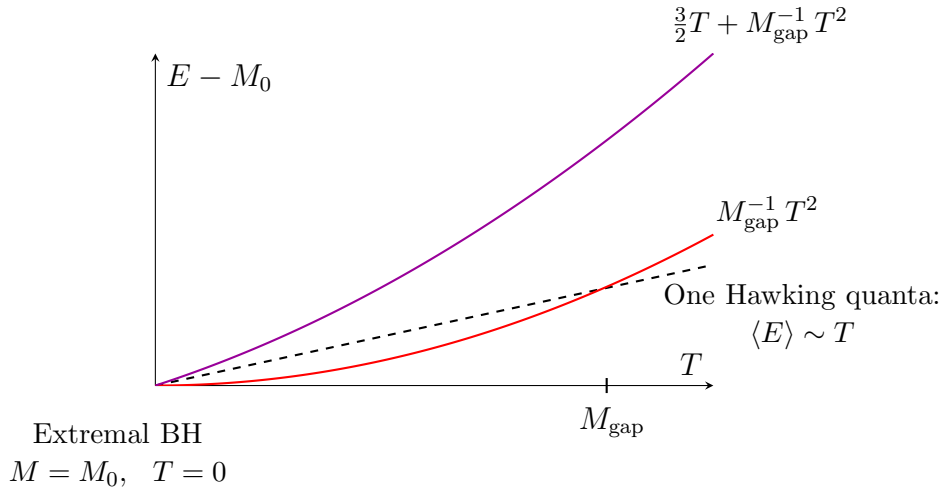
\begin{figure}[t!]
\begin{center}
  \begin{tikzpicture}[scale=1]
  \begin{axis}[
      width=9cm,
      height=6cm,
      xtick=\empty, 
      ytick=\empty, 
      xlabel=$T$, 
      ylabel=$E-M_\text{0}$,
      axis lines=center, 
      domain=0:1.25,
      samples=100]
        \addplot [thick, dashed] {x};
    \addplot [red, thick] {x^2};
     \addplot[purple, thick]{ (3*x/2)+x^2};
  \end{axis}
   \node[text width=6cm] at (-0.5,-0.82) {\begin{center}{\small Extremal BH \\ $M=M_0$, \, $T=0$}\end{center}};
    \node[text width=6cm] at (8.6-0.4,2.55) {\begin{center}{\small $M_\text{gap}^{-1}\, T^2\,\,$ }\end{center}};
        \node[text width=6cm] at (8.6-0.5,5.06) {\begin{center}{\small $\frac{3}2T+  M_\text{gap}^{-1}\, T^2\,\,$ }\end{center}};
    
        \node[text width=6cm] at (8.6,1.2) {\begin{center}{\small One Hawking quanta:\\
$
\<E\> \sim T $ }\end{center}};
    \draw[style=thick, ] {(6.0,-0.1)  -- (6.0,0.1)}; 
        \node[text width=6cm] at (6.0,-0.2) {\begin{center}{$~M_\text{gap}$ }\end{center}};
  \end{tikzpicture}
  \caption{\label{fig:plot-of-energy-deivation}Energy above extremality at fixed charge as a function of the temperature when obtained from the semiclassical analysis (in red) and when accounting for the quantum fluctuations in the near-horizon region (in purple). This should be compared to the average energy of one Hawking quanta (dashed line) whose energy is on average $\< E\> \sim T$.}
    \end{center}
  \end{figure}
  
\begin{figure}[t!]
\begin{center}
 \hspace{-3.7cm} \begin{tikzpicture}[scale=1]
  \begin{axis}[
      width=12cm,
      height=7cm,
      xtick=\empty, 
      ytick=\empty, 
      xlabel=$E-M_0$, 
      ylabel=$\rho(E)$,
      axis lines=center, 
      domain=0:1.25,
      samples=100]
     \addplot[red, thick, samples=500]{ 1/2*e^(2*((x/1.2)^(1/2)))};
        \addplot [purple, thick, samples=500] {sinh(2*((x/1.2)^(1/2)))};
  \end{axis}
     \draw[style=thick] {(3.0,-0.1)  -- (3.0,0.1)}; 
        \node[text width=6cm] at (3.0,-0.2) {\begin{center}{$~M_\text{gap}$ }\end{center}};
            \draw[style=thick] {(-0.1,0.71)  -- (0.1,0.71)}; 
            \node[text width=6cm] at (-0.65,1.1) {\begin{center}{$~e^{\frac{A_{0}}{4G_N}}$ }\end{center}};
  \end{tikzpicture}
  \caption{\label{fig:plot-of-rho} Purple: Density of states (at fixed charge) for black holes states as a function of energy above extremality $E-M_0$, including backreaction effects given in \eqref{eq:dossinh}. Red: Plot of the naive density of states $\rho\sim \exp{(A_{\rm hor}/4 G_N)}$ which starts deviating from the full answer below energies of order $M_{\rm gap}$.}
    \end{center}
  \end{figure}
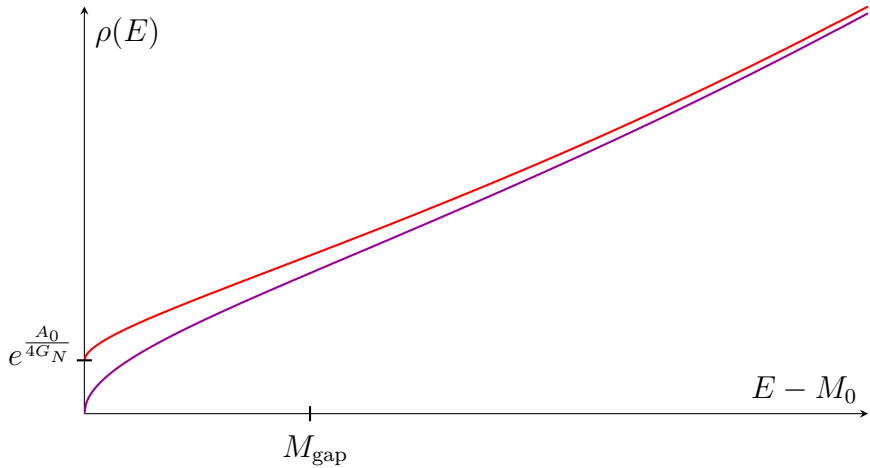

In this paper, we settle the debate about the existence of a mass-gap for $4d$ Reissner-Nordstr\"{o}m black holes. We show that such near-extremal black holes, in fact, do not exhibit a mass gap at the scale $M_{\text{gap}}$.\footnote{While in this paper we will mostly focus on studying $4d$ black holes in an asymptotically flat or $AdS_4$ space, our analysis could be applied to RN black holes in any number of dimensions.  }  To arrive at this conclusion, we go beyond the semiclassical analysis and account for quantum fluctuations to reliably compute the partition function of such black holes at temperatures $T \sim M_\text{gap}$ in the canonical and grand canonical ensembles. By taking the Laplace transform of the partition function, we find the density of states in the spectrum of black holes masses. Due to the presence of $T \log (T/M_{\rm gap})$ corrections to the free energy, \footnote{The $ T \log T$ corrections discussed throughout this paper should not be confused with the logarithmic area corrections to the entropy studied for extremal black holes in \cite{Banerjee:2010qc, Banerjee:2011jp, Sen:2011ba, Sen:2012cj}. While we did not find any connection between the two corrections (as the logarithmic area correction to the entropy is studied in a specific limit for the mass, charge and temperature; such a limit is not employed in this paper), it would be interesting to understand whether the results obtained in this paper can be used to also account for the entropy corrections from \cite{Banerjee:2010qc, Banerjee:2011jp, Sen:2011ba, Sen:2012cj}. It should also not be confused with logarithmic corrections coming from a saddle-point evaluation of a Laplace transform such as in \cite{Loran:2010bd}.}  we find that the spectrum looks like a continuum of states and, consequently, exhibits no gap of order $\sim M_{\text{gap}}$.  This continuum is observed because our computation is not sensitive enough to distinguish between individual black hole microstates; for that, an ultraviolet completion of the gravitational theory is necessary. Nevertheless, our computation does suggest that, for non-supersymmetric theories, the degeneracy of extremal black holes is much smaller than that obtained from the area-law Bekenstein-Hawking entropy (in figure \ref{fig:plot-of-rho} we show the shape of the density of states at fixed charge)\footnote{The logic in this paper is very different from the argument in \cite{Hawking:1994ii}. The degeneracy of the extremal black hole and the presence of a gap depends on the amount of supersymmetry in the theory (see section \ref{sec:black-holes-with-diff-matter}).}.

The potential breakdown of Hawking's analysis raised in \cite{Preskill:1991tb} is also resolved. In figure \ref{fig:plot-of-energy-deivation}, we compare the temperature dependence of the energy above extremality in the classical analysis and when accounting for quantum fluctuations. We find that when only slightly above extremality, $E - M_0 \sim \frac{3}2 T>T$, therefore resolving the naive failure of thermodynamics at very small temperatures. Nevertheless, we should stress that the we expect the spectrum of Hawking radiation to be strongly modified at such low temperatures.  

 A similar analysis was done recently for the case of near-extremal rotating BTZ black holes in AdS$_3$ \cite{Ghosh:2019rcj}. These black holes also present a breakdown of their statistical description at low temperatures when restricted to the semiclassical analysis. The breakdown is similarly resolved by including backreaction effects in the Euclidean path integral. 

To reliably compute the partition function at such small temperatures, we perform a dimensional reduction to the two dimensional $AdS_2$ space in the near-horizon region\footnote{The geometry describing the throat is AdS$_2\times$S${}^2$. Even though the size of the transverse sphere $r_0$ is large, we will consider temperatures well below the KK scale $T\ll M_{\rm KK} \sim 1/r_0$. This is consistent since, in all cases, we study the gap is a parametrically smaller scale $T\sim M_{\rm gap} \ll M_{\rm KK}$.}. We find that the only relevant degrees of freedom that affect the density of states are the massless modes coming from the gravitational sector, the electromagnetic gauge field, and the $SO(3)$ gauge fields generated by the dimensional reduction. The resulting effective theory turns out to be that of $2d$ Jackiw-Teitelboim (JT) gravity \cite{Teitelboim:1983ux, Jackiw:1984je} coupled to gauge degrees of freedom. The Euclidean path integral of such an effective theory can be computed exactly by first integrating out the gauge degrees \cite{Iliesiu:2019lfc, Kapec:2019ecr} and then by analyzing the boundary modes \cite{Almheiri:2014cka} of the resulting model using the well-studied Schwarzian theory \cite{kitaevTalks, Maldacena:2016hyu, Jensen:2016pah, Maldacena:2016upp, Engelsoy:2016xyb, Jevicki:2016ito, Bagrets:2016cdf, Stanford:2017thb, Mertens:2017mtv, Lam:2018pvp, Goel:2018ubv, Kitaev:2018wpr, Yang:2018gdb, Saad:2019lba, Mertens:2019tcm, Iliesiu:2019xuh}. 

The connection between JT gravity and near-extremal black holes has been widely discussed in past literature \cite{Almheiri:2016fws, Anninos:2017cnw, Sarosi:2017ykf, Nayak:2018qej, Moitra:2018jqs, Hadar:2018izi, Castro:2018ffi, Larsen:2018cts,  Moitra:2019bub, Sachdev:2019bjn, Hong:2019tsx, Castro:2019crn, Charles:2019tiu}. In fact, in \cite{Almheiri:2016fws}, the scale $M_{\text{gap}}$ defined through the thermodynamics was identified as the symmetry breaking scale for the emergent near-horizon AdS$_2$ isometries, $SL(2,\mR)$. Moreover, this is also the scale at which the equivalent Schwarzian theory becomes strongly coupled. However, compared to past literature, to compute the partition function at small temperatures, $T\sim  M_{\text{gap}}$,  we had to keep track of all the fields generated through the dimensional reduction and exactly compute the path integral for the remaining massless relevant degrees of freedom. Our qualitative picture is nevertheless similar to that presented in \cite{Almheiri:2016fws} as we show that the semiclassical analysis fails due to the backreaction of the dilaton and gauge fields on the metric. 

For the reasons described above, to avoid confusion from now on, we will stop calling the scale in which the semiclassical analysis breaks down $M_{\rm gap}$ since there is no gap at that scale. Instead we will redefine it as $M_{\rm gap} \to \frac{1}{2 \pi^2} M_{{\text{\tiny{SL(2)}}}}$. The factor of $2\pi^2$ will be useful but is just conventional. More importantly, we want to stress that the appropriate meaning of this energy is really the symmetry breaking scale of the approximate near horizon conformal symmetry. 

 The rest of this paper is organized as follows. In section \ref{sec:near-extremal-in-AdS4}, we describe the set-up for Reissner-Nordstr\"{o}m black holes, discuss details about the dimensional reduction, dynamics and boundary conditions for massless fields in the near-horizon region. In section \ref{sec:part-function-large-BHs-AdS4}, we reduce the dynamics in the near-horizon region to that of a $1d$ system, the Schwarzian theory coupled to a particle moving on a $U(1)\times SO(3)$ group manifold.  We compute the partition function and density of states in such a system in the canonical and grand canonical ensembles, thus obtaining the main result of this paper in section \ref{sec:large-BHs-AdS4-canonical-ensemble} and \ref{sec:DOS}. In section \ref{sec:large-BHs-AdS4-grand-canonical-ensemble-2}, we also account for deviations from the spinless Reissner-Nordstr\"{o}m solution to Kerr-Newman solutions with small spin, in a grand canonical ensemble that includes a chemical potential for the angular momentum (or equivalently, fixing the boundary metric). More details about the connection between the $SO(3)$ gauge field appearing from the dimensional reduction and the angular momentum of the black hole are discussed in appendix \ref{app:SO(3)-gauge-field-review}. In section \ref{sec:massive-KK-modes}, we revisit the contribution of massive Kaluza-Klein modes to the partition function. We show their effect is minimal and does not modify the shape of the density of states. Finally, in section~\ref{sec:conclusion} we summarize our results and discuss future research directions,  focusing on possible non-perturbative corrections to the partition function and speculating about the role that geometries with higher topology have in the near-horizon region.

\section{Near-extremal black hole and JT gravity}
\label{sec:near-extremal-in-AdS4}

In this paper, we will focus on several kinds of $4d$ black hole solutions. Specifically, in this section, we will consider the Reissner-Nordstr\"{o}m  black holes solutions and Kerr-Newman solutions of low spin, in both asymptotically $AdS_4$ spaces and flat spaces. 
While here we focus on black holes in $D=4$, the techniques used here apply to a broader set of near-extremal black holes in any number of dimensions. 

\subsection{Setup}
\label{sec:setup}

In this section we will study Einstein gravity in asymptotically AdS$_4$ coupled to a $U(1)$ Maxwell field. The Euclidean action is given by 
\be 
\label{eq:Einstein-Maxwell-action}
I_{EM} =  &-\frac{1}{16\pi G_N}\left[ \int_{M_4}  d^4 x \sqrt{g_{(M_4)}} \left(R +2 \Lambda \right) - 2 \int_{\partial M_4} \sqrt{h_{\partial M_4}} K\right] \nn \\ &-\frac{1}{4 e^2} \int_{M_4} d^4x \sqrt{g_{(M_4)}} \,F_{\mu \nu} F^{\mu \nu} \,,
\ee
where $F = dA$ and where we take $A$ to be purely imaginary.  
The coupling constant of the gauge field is given by $e$, and $\Lambda=3/L^2$ denotes the cosmological constant with corresponding AdS radius $L$. It will be more intuitive to sometimes keep track of $G_N$ by using the Planck length instead, $G_N = \lpl^2$. 

The focus of this paper will be to compute the Euclidean path integral (fixing boundary conditions in the boundary of flat space or $AdS_4$) around certain background geometries. Throughout this paper, we fix the boundary metric $h_{ij}$ of the manifold $M_4$, which requires the addition of the Gibbons-Hawking-York term in \eqref{eq:Einstein-Maxwell-action}. For the gauge field, we will pick boundary conditions dominated by solutions with a large charge at low temperatures, in the regime where the black hole will be close to extremality. Specifically, the two boundary conditions that we will study will be: 
\begin{itemize}
\item Fixing the components of $A_i$ along the boundary $\partial M_4$. With such boundary conditions, \eqref{eq:Einstein-Maxwell-action} is a well defined variational problem.  As we will see shortly,  dimensionally reducing the action \eqref{eq:Einstein-Maxwell-action} to $2d$, amounts to fixing the holonomy around the black hole's thermal circle; in turn, this amounts to studying the system in the charge grand canonical ensemble with the holonomy identified as a chemical potential for the black hole's charge.

\item We will also be interested in fixing the charge of the black hole, which corresponds to studying the charge microcanonical ensemble. Fixing the charge amounts to fixing the field strength $F_{ij}$ on the boundary. In this case, we need to add an extra boundary term for \eqref{eq:Einstein-Maxwell-action} to have a well defined variational principle \cite{Braden:1990hw, Hawking:1995ap}
\be
\label{eq:boundary-term-fixed-F}
\tilde{I}_{EM} = I_{EM} - \frac{1}{e^2} \int_{\partial M_4} \sqrt{h} F^{ij} \,\hat n_{i} A_{j} \,,
\ee
where $\hat n$ is outwards unit vector normal to the boundary. To compute the free energy in the case of black holes in AdS$_4$, we could alternatively add the usual holographic counterterms in the AdS$_4$ boundary \cite{Henningson:1998ey,Balasubramanian:1999re}. A detailed analysis of all possible saddles was done in \cite{Chamblin:1999tk}. For our purposes, we will focus on the charged black hole contribution.
\end{itemize}

To start, we review the classical  Reissner-Nordstr\"{o}m solution of \eqref{eq:Einstein-Maxwell-action}, obtained when fixing the field strength on the boundary and consequently the overall charge of the system. The metric is given by 
\be \label{eq:RN-black-hole-AdS}
ds^2_{(4d)} &= f(r) d\tau^2 + \frac{dr^2}{f(r)} +r^2 d\Omega_2^2\,,~~~~~f(r)=1- \frac{2G_N M}{r} +\frac{G_N}{4 \pi} \frac{Q^2}{r^2} + \frac{r^2}{L^2}\,,
\ee
For concreteness we will pick the pure electric solution with $F= {\small \frac{e Q}{4\pi}} * \hspace{-0.1mm}\epsilon_2$, with $\epsilon_2$ the volume form on $S^2$, while the magnetic solution has $F={\small \frac{e Q}{4\pi}}  \epsilon_2$.\footnote{As we will show shortly, in this units the charge is quantized as $Q \in e \cdot \mZ$.} Such black holes have two horizons $r_+$ and $r_-$ located at the zeroes of $f(r_\pm)=0$. We will refer to the larger solution as the actual horizon radius $r_h=r_+$. As a function of the charge, the temperature and chemical potential are given by 
\be\label{rnbmu}
\beta = \frac{4\pi}{|f'(r_h)|} ,~~~~ \mu = \frac{e}{4\pi} \frac{Q}{r_h}\,.
\ee
In terms of the chemical potential the vector potential can be written as $A =i \mu\left( 1- \frac{r_h}{r}\right) d\tau$ such that its holonomy is $e^{\mu \beta}$ along the boundary thermal circle. The Bekenstein-Hawking entropy for these black holes is given by 
\be
S = \frac{A}{4 G_N} = \frac{\pi r_h^2}{G_N}\,.
\ee
However, as we will see below, if the entropy is defined through the Gibbons-Hawking procedure instead, the result can be very different due to large fluctuations in the metric. To enhance this effects we will consider the regime of low temperatures and large charge next.    

\begin{center} 
\textbf{Near-extremal Limits}
\end{center} 

 In the extremal limit, both radii become degenerate and $f(r)$ develops a double zero at $r_0$ (which can be written in terms of for example the charge). In this casem the extremal mass, charge and Bekenstein-Hawking entropy are given by 
\be\label{extpar}
Q^2 =\frac{4\pi }{G_N} \left(r_0^2 + \frac{3r_0^4}{L^2}\right) ,~~~~M_0 = \frac{r_0}{G_N}\left( 1 + \frac{2r_0^2}{L^2}\right),~~~~S_0 = \frac{\pi r_0^2}{G_N}\,.
\ee 
This is the naive zero temperature extremal black hole. As we will see below, the small temperature limit of the entropy will not be given by the extremal area $S_0$ but it will still be a useful parameter to keep track of.

Since the semiclassical description breaks down at sufficiently small temperatures, we will study near-extremal large black holes with very large $\beta = T^{-1}$. We will first review its semiclassical thermodynamics in this limit. To be concrete, we will do it here by fixing the charge and the temperature. We will write the horizon radius as $r_h = r_0 +\delta r_h$ where $r_0$ is the extremal size for the given charge. Then the temperature is related to $\delta r_h$ as 
\beq
r_h = r_0 +\delta r_h,~~~~~\delta r_h = \frac{2\pi}{\beta}L_2^2 + \ldots ,~~~~L_2\equiv \frac{L r_0}{\sqrt{L^2+6 r_0^2}}\,,
\eeq
where the dots denote sub-leading terms in the large $\beta$ limit and the physical interpretation of the quantity $L_2(r_0)$ will become clear later. The energy and Bekenstein-Hawking entropy if we fix the charge behave as 
\be
E(\beta,Q)=M_0 + \frac{2\pi^2}{M_{{\text{\tiny{SL(2)}}}}}T^2 + \ldots  ,~~~~~S(\beta,Q)=S_0 + \frac{4\pi^2}{M_{{\text{\tiny{SL(2)}}}}}T+\ldots, 
\ee
where the dots denote terms suppressed at low temperatures, and where we define the gap scale
\be
M_{{\text{\tiny{SL(2)}}}}^{-1} \equiv \frac{r_0 L_2^2}{G_N} ,
\ee
where $r_0$ is a function of the charge given by \eqref{eq:extremalparam}. Due to this scaling with temperature, as reviewed in the introduction, the statistical description breaks down at low temperatures $\beta \gtrsim M^{-1}_{{\text{\tiny{SL(2)}}}}$ so we identify this parameter with the proposed gap scale of \cite{Preskill:1991tb} (as anticipated in the introduction, we will see in the next section that this intuition is wrong). A similar analysis to the one above can be done for fixed chemical potential. 

Two limits of this near-extremal black hole will be particularly useful. The first is the limit $L\to \infty$ where we recover a near-extremal black hole in flat space, and large $Q$. In this case the mass and entropy scale with the charge as 
\beq
r_0 \sim  \ell_{\rm PL} Q,~~~M_0 \sim  \frac{Q}{\ell_{\rm PL}},~~~S_0 \sim Q^2. 
\eeq
We will take the limit also of large charge $Q$ for two reasons. First, we want the black hole to be macroscopic with a large size compared with Planck's length. Second, we want $S_0 \gg1$. As we will see below, this will suppress topology changing processes near the horizon \cite{Saad:2019lba}. In this limit $M_{{\text{\tiny{SL(2)}}}} \sim G_N/Q^3$. 

The second limit we will consider is a large black hole in AdS, keeping $L$ fixed. Following \cite{Nayak:2018qej} we will take large charges such that $r_0 \gg L$. We achieve this by choosing boundary conditions such that $Q \gg L/\ell_{Pl}$ (or $\mu \gg e / \ell_{Pl}$). In this regime the charge and mass are approximately 
\be\label{eq:extremalparam}
Q^2 = \frac{4\pi }{G_N}\frac{3r_0^4}{L^2} ,~~~~~M_{0}= \frac{2r_0^3}{G_NL^2} \sim Q^{3/2} ,~~~~~S_0 =  \frac{\pi r_0^2}{G_N}\sim Q\,.
\ee 
 For a bulk of dimension $D=d+1$, the mass of the extremal state scales as $M_0 \sim Q^{\frac{d}{d-1}}$ for large charge. This scaling is dual to the thermodynamic limit of the boundary CFT$_d$ in a state with finite energy and charge density, see for example \cite{Jafferis:2017zna}. Since $L\gg \ell_{Pl}$, then $r_0 \gg L$ implies $r_0 \gg \ell_{Pl}$ and therefore $S_0 \gg 1$, suppressing topology changing processes near the horizon. In this limit $M_{{\text{\tiny{SL(2)}}}} \sim G_N^{3/4}/Q^{1/2}$.

\begin{center}
\textbf{Near-extremal Geometry}
\end{center}

Finally, in the near-extremal limit we will divide the bulk geometry in a physically sensible way that will be very useful below \cite{Nayak:2018qej}. We will separately analyze the near-horizon region and the far region, as depicted in figure \ref{fig:regions}. They are described as: 
\begin{figure}[t!]
\begin{center}
\begin{tikzpicture}
\draw[purple,  thick]  (0,-1.0) -- (0,1.0);
\node[text width=2cm] at (1.1,0) {\small Horizon};
 \draw[red, thick] (0, -1.0) .. controls (7,-1.2) .. (10, -3.5);
 \node[text width=2cm] at (4,2.2) {\begin{center}{ \textit{NHR} \\ AdS$_2 \times S^2$ \\
 JT gravity }\end{center}};
  \node[text width=4.0cm] at (3.8,0) {\begin{center}{\small Quantization is
 easy\\ (linear dilaton)}\end{center}};
  \node[text width=3.5cm] at (9.8, 1.9) {\begin{center}{\textit{FAR} \\ Near-extremal solution}\end{center} };
   \node[text width=4cm] at (9.8,0) {\begin{center}{\small }\end{center}};
  \draw[red, thick] (0, 1.0) .. controls (7,1.2) .. (10, 3.5);
   \draw[blue,thick] plot [smooth,tension=0.6] coordinates {(6,-1.35)(6.2,-0.5) (6.15,-0.25) (6, 0)(5.8,0.5)(6,1.35)};
\end{tikzpicture}
\end{center}
\caption{A cartoon of the near-horizon region (NHR) and the far-away region (FAR) separated by a boundary at which the boundary term of JT gravity will need to be evaluated. In the throat quantization is easy and necessary to account for at low temperatures. In the FAR quantization is hard but quantum corrections are suppressed.}\label{fig:regions}
\end{figure}
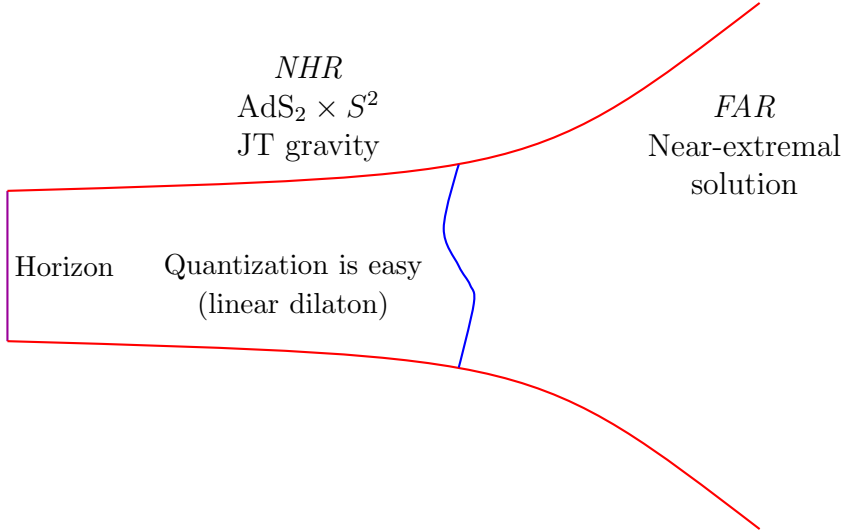

\textbf{Near-horizon region (NHR):} This is located at radial distances $r-r_0 \ll r_0$ and is approximately AdS$_2 \times S^2$ with an AdS$_2$ and $S^2$ radius given by
\beq
L_2=\frac{L r_0}{\sqrt{L^2+6 r_0^2}},~~R_{S^2}=r_0.
\eeq
Indeed from the metric \eqref{eq:RN-black-hole-AdS} we can approximate, defining $\rho = r-r_0$, in the near-horizon region
\beq
ds^2_{(4d)} = \frac{\rho^2 - \delta r^2_h}{L_2^2} d\tau^2 + \frac{L_2^2}{\rho^2 - \delta r^2_h} d\rho^2 + (r_0 + \rho)^2 d\Omega_2
\eeq
where the first two terms correspond to the thermal AdS$_2$ factor with AdS radius $L_2 $ and the second factor is a sphere with an approximately constant radius $r_0$. For a black hole in flat space limit the radius of AdS$_2$ is $L_2 \approx r_0$ while for a large black hole in AdS it is given by $L_2 \approx L/\sqrt{6}$.

We kept the slowly varying term in the size of the transverse $S^2$ since this small correction breaks the AdS$_2$ symmetries and dominates the low-temperature dynamics \cite{Almheiri:2014cka, Maldacena:2016upp}. As indicated in figure \ref{fig:regions}, we will review how the four-dimensional theory reduces to JT gravity in this region. At positions $\rho \gg \delta r_h$, the finite temperature effects can be neglected, and the geometry becomes vacuum AdS$_2$. This condition also guarantees that we are near the conformal boundary of AdS$_2$. Since we will take very low temperatures $\delta r_h \ll L_2$ and therefore it is also true that $\delta r_h \ll r_0$ \footnote{\textbf{v3:} In a previous version of this paper it was incorrectly stated that being close to the conformal boundary of AdS$_2$ requires $\rho \gg L_2$, while the correct inequality is $\rho \gg \delta r_h$. The previous version would lead to the incorrect conclusion that the gluing procedure used here is not valid in flat space.}.

We also look at the behavior of the $U(1)$ field strength in this region $F_{\tau \rho} \sim Q/(4\pi r_0^2)$. Therefore the throat is supported by a constant electric field.

\textbf{Far-away region (FAR):}  This is located instead at large $r$, where the metric can be approximated by the extremal AdS$_4$ metric 
\beq
\label{eq:extremal-metric}
ds^2_{(4d)} = f_0(r) d\tau^2 + \frac{dr^2}{f_0(r)} + r^2 d\Omega_2,~~~f_0(r) = \frac{(r-r_0)^2}{r^2 L^2} (L^2 + 3 r_0^2 + 2 r r_0 + r^2)
\eeq
with the identification $\tau \sim \tau + \beta$. As the temperature is taken to zero this region keeps being well approximated by the semiclassical geometry. This is appropriate for the case of large black hole limit in AdS$_4$. For the case of black holes in the flat space limit, we take $L\to\infty$ of the metric above, finding the extremal geometry in asymptotically flat space.

Both the NHR and the FAR region overlap inside the bulk. We will match the calculations in each region at a surface included in the overlap, denoted by the blue line in figure \ref{fig:regions}. This happens at radial distances such that $\delta r_h \ll r-r_0 \ll r_0$. We will denote the gluing radius by $r_{\partial M_{\rm NHR}} = r_0 + \delta r_\bdy$, but as we will see below, the leading low-temperature effects are independent of the particular choice of $r_{\partial M_{\rm NHR}}$ as long as its part of the overlapping region.

\subsection{Dimensional reduction}
\label{sec:dimensional-reduction}

So far, we analyzed the semiclassical limit of large near-extremal black holes. We explained how the full four-dimensional geometry decomposes in two regions near the horizon throat (NHR) and far from the horizon (FAR). The parameter controlling quantum effects in the FAR region is $G_N$ which we always keep small, while in the throat the parameter becomes the inverse temperature $\beta M_{{\text{\tiny{SL(2)}}}} $ (due to the pattern of symmetry breaking). Since the geometry in the throat is nearly AdS$_2 \times S^2$ we can do a KK reduction on the transverse sphere, and the dominant effects become effectively two dimensional.  

In this section, we will work out the dimensional reduction from four dimensions to two dimensions. With respect to \cite{Nayak:2018qej}, our new ingredients will be to point out that the reduction works for low temperatures $\beta M_{{\text{\tiny{SL(2)}}}} \gtrsim 1$ where the semiclassical approximation breaks down, and to include the $SO(3)$ gauge mode associated to diffeomorphisms of the transverse sphere. 
We will begin by analyzing the reduction of the metric and will include the gauge fields afterwards. The ansatz for the four dimensional metric that we will use, following \cite{Salam:1984cj}, is
\be
\label{eq:beyond-S-wave-metric-ansatz}
ds^2_{(4d)}  =  \frac{r_0}{\chi^{1/2}} g_{\mu\nu}dx^\mu dx^\nu+ \chi ~ h_{mn} (dy^m + \mathbf{B}^a \xi^m_a  )(dy^n + \mathbf{B}^b \xi^n_b )\,,
\ee
where $x^\mu=(\tau,\rho)$ label coordinates on $AdS_2$ and $y^m=(\theta,\phi)$ coordinates on $S^2$ with metric $h_{mn}={\rm diag}(1,\sin^2\theta)$. At this point $r_0$ is a constant parameter which will later be chosen to coincide with the extremal radius introduced above, when we look at solutions. The size of the transverse sphere is parametrized by the dilaton $\chi(x)$ while we also include the remaining massless mode from sphere fluctuations $\mathbf{B}$. We can use diffeomorphisms to make the gauge field independent of the coordinates on $S^2$, so $\mathbf{B}^a = B^a_\mu(x) dx^\mu$. Here $\xi_a = \xi^n_a \partial_n$ are the (three) Killing vectors on $S^2$ given by
\be
\xi_1 &= \cos \varphi \partial_\theta - \cot \theta \sin \varphi \partial_\varphi, \nn \\ 
\xi_2 &=- \sin \varphi \partial_\theta - \cot \theta \cos \varphi \partial_\varphi, \nn \\
\xi_3 &= \partial_\varphi,
\ee
and via the Lie bracket $[\xi_a, \xi_b] =\varepsilon_{abc} \xi_c$ they generate the Lie algebra of the $SU(2)$ isometry group. The consistency of this reduction was analyzed perturbatively in \cite{Michelson:1999kn}. Some useful technical results involving this ansatz were derived in \cite{GibbonsPope}. The Einstein action after the reduction, keeping only massless fields, is
\be
I^{(2d)}_{EH}=& -\frac{1}{4G_N}\left[ \int_{M_2} d^2 x \sqrt{g}  [\chi R - 2U(\chi)] +2 \int_{\partial M_2} du \sqrt{h} \chi K \right] \nn\\
&- \frac{1}{12G_N r_0} \int_{M_2} d^2 x   \sqrt{g} \chi^{5/2} ~{\rm Tr}(H_{\mu\nu}H^{\mu\nu})\,,
\ee
which has the form of a two dimensional dilaton-gravity theory coupled to $SO(3)$ Yang-Mills field with dilaton potential and field strength
\beq
U(\chi) = -r_0 \left( \frac{3 \chi^{1/2}}{L^2} + \frac{1}{\chi^{1/2}} \right),
\eeq
We also defined a $SO(3)$ valued field $B = B_\mu ^a T^a dx^\mu$, with $T^a$ antihermitian generators in the adjoint representation normalized such that $[T^a,T^b] =\varepsilon_{abc} T^c$ and ${\rm Tr} ( T^aT^b) = -{\small \frac{1}{2}} \delta^{ab}$, and field strength $H= dB -B \wedge B$. We will see below how in the state corresponding to a large near-extremal black hole this reduces to Jackiw-Teitelboim gravity \cite{Jackiw:1984je,Teitelboim:1983ux}.

Finally, we can reduce the Maxwell term to the massless s-wave sector. In order to do this, we decompose the gauge field as \cite{Michelson:1999kn}\footnote{The expansion in  \eqref{eq:A-expansion} assumes that no overall magnetic flux is thread through $S^2$. }
\bea
\label{eq:A-expansion}
A_\mu(x,y) &=& a_\mu (x) \frac{1}{\sqrt{4\pi}}+ \sum_{\ell\geq 1,m} a_\mu^{(\ell,m)}(x) Y_\ell^m(y), \\ 
A_n(x,y) &=& \sum_{\ell\geq 1,m} a^{(\ell,m)}(x) \epsilon_{np} \nabla^pY_\ell^m(y) + \sum_{\ell\geq 1,m}  \tilde{a}^{(\ell,m)}(x) \nabla_n Y_\ell^m(y),
\ea
where in the first line $Y_\ell^m(y)$ are the scalar spherical harmonics in $S^2$, and in the second line we wrote the vector spherical harmonics in terms of the scalar ones. This decomposition shows that the only massless field after reduction is the two dimensional s-wave gauge field $a_\mu(x)$. In the second line we see there is no component for $A_n$ that is constant on $S^2$ (since such configurations would yield a singular contribution to the action from the poles of $S^2$) and therefore no other massless field is generated. Therefore the s-wave massless sector of the Maxwell action becomes 
\beq
I^{(2d)}_M = - \frac{1}{4e^2 r_0 } \int_{M_2} d^2 x  \sqrt{g} \chi^{3/2}  f_{\mu\nu}f^{\mu\nu},~~~~f=da
\eeq

Putting everything together, the massless sector  of the dimensionally reduced Einstein-Maxwell action \eqref{eq:Einstein-Maxwell-action}  is given by 
\bea
\label{eq:action-dimensionally-reduced-final}
I^{(2d)}_{EM}&=& -\frac{1}{4G_N}\left[ \int_{M_2}d^2 x \sqrt{g}  [\chi R -2 U(\chi)] +2 \int_{\partial M_2} du \sqrt{h} \chi K \right]\nn\\
&& - \frac{1}{12G_N r_0} \int_{M_2} d^2x \sqrt{g} \chi^{5/2} ~{\rm Tr}(H_{\mu\nu}H^{\mu\nu}) - \frac{1}{4e^2 r_0} \int_{M_2} d^2 x \sqrt{g} \chi^{3/2} f_{\mu\nu}f^{\mu\nu}\,,
\ea
where the first terms corresponds to two dimensional gravity, the second to the $SO(3)$ gauge theory generated from the KK reduction and the third to the reduction of the four dimensional $U(1)$ gauge field. The contribution of the remaining massive fields coming from the $U(1)$ gauge field, metric or other potential matter couplings is summarized in section \ref{sec:massive-KK-modes} and their contribution to the partition function is discussed in section \ref{sec:KKmodespart}. As explained in the introduction, such modes are shown to have a suppressed contribution at low temperatures and, therefore, in order to answer whether or not there is an energy gap for near-extremal black holes it is sufficient to study the contribution of the massless fields from \eqref{eq:action-dimensionally-reduced-final}.  Consequently, we proceed by studying the quantization of the $2d$ gauge field in \eqref{eq:action-dimensionally-reduced-final}, neglecting the coupling of the $SO(3)$ gauge field to the massive Kaluza-Klein modes and coupling of the $U(1)$ gauge field to other potential matter fields that can be present in \eqref{eq:Einstein-Maxwell-action}.

\subsection{Two dimensional gauge fields} 
\label{sec:two-dimensional-gauge-fields}
In order to proceed with the quantization of the gauge field in \eqref{eq:action-dimensionally-reduced-final} it is necessary to introduce two Lagrange multipliers zero-form fields, $\phi^{U(1)}$ and $\phi^{SO(3)}$, with the latter valued in the adjoint representation of $SO(3)$. The path integral over the gauge fields with action  \eqref{eq:action-dimensionally-reduced-final} can be related to the path integral over $A$, $B$ and $\phi^{U(1),\,SO(3)}$ for the action
\be
\label{eq:action-dimensionally-reduced-final-w-Lagr-multiplier}
\tilde I_{EM}&= -\frac{1}{4G_N}\left[ \int_{M_2} d^2 x \sqrt{g} [\chi R - 2U(\chi)] +2 \int_{\partial M_2} du \sqrt{h} \chi K \right]\nn\\
&\hspace{-0.5cm}- i \int_{M_2} \left(\phi^{U(1)} f + \tr \phi^{SO(3)}H\right)-   \int_{M_2} d^2 x \sqrt{g} \left[ \frac{3G_N r_0}{2\chi^{5/2} }~{\rm Tr}(\phi^{SO(3)})^2 + \frac{e^2 r_0}{2\chi^{3/2}}  (\phi^{U(1)})^2\right]\,,
\ee
by integrating out the Lagrange multipliers $\phi^{U(1), \,SO(3)}$. One subtlety arises in going between \eqref{eq:action-dimensionally-reduced-final-w-Lagr-multiplier} and \eqref{eq:action-dimensionally-reduced-final}. When integrating-out $\phi^{U(1),\, SO(3)}$ there is a one-loop determinant which depends on the dilaton field $\chi$ which yields a divergent contribution to the  measure (behaving as $\exp\, 4 \delta(0) \int_{M_2} du \log \chi(u) $) for the remaining dilaton path integral. There are two possible resolutions to this problem. The first is to define the measure for the dilaton path integral for the action \eqref{eq:action-dimensionally-reduced-final} in such a way that it cancels the contribution of the one-loop determinant coming from \eqref{eq:action-dimensionally-reduced-final-w-Lagr-multiplier} \footnote{\textbf{v2:} This definition of the measure which removes the dilaton dependence of the one-loop determinant precisely arises when derived from the higher dimensional path integral measure. In the context of 3D gravity, see for example footnote 4 of \cite{Maxfield:2020ale} (published version).}. The second resolution is to rely on the fact that logarithmic corrections to the free energy (that are of interest in this paper)  solely come from integrating out fields in the near-horizon region. However, as we will see shortly, in the near-horizon region, the dilaton field $\chi$ is dominated by its value at the horizon and consequently the one-loop determinant is simply a divergent constant which can be removed by the addition of counterterms to the initial action \eqref{eq:action-dimensionally-reduced-final}. Regardless of which resolution we implement, the gauge degrees of freedom in two dimensional Yang-Mills theory coupled to dilaton gravity as in \eqref{eq:action-dimensionally-reduced-final-w-Lagr-multiplier} can be easily integrated-out \cite{Iliesiu:2019lfc}. 

To begin, we fix the gauge field along the three-dimensional boundary which implies that we are also fixing the holonomy at the boundary $\partial M_2$, $e^{\mu} = \exp \oint a$ and take $e^{i \beta \mu_{\text{\tiny{SO(3)}}} \sigma_3} \sim [\cP \exp (\oint B)]$.\footnote{Here, and throughout the rest of this paper, $\sim$ specifies equality of conjugacy classes. The meaning of the holonomy for the $SO(3)$ gauge field arising from the dimensional reduction is that as one observer travels along $\partial M_2$ the internal space $S^2$ is rotated by an angle $\mu_{\text{\tiny{SO(3)}}}$ around a given axis.   } In such a case we find that by integrating out the gauge degrees of freedom yields an effective theory of dilaton gravity for each $U(1)$ charge $Q$ and each $SO(3)$ representation $j$: 
\beq
\label{eq:part-function-sum-over-Q,J}
Z_{\rm RN}[\mu,\beta] = \sum_{Q\in e \cdot \mZ,~j\in \mZ} (2j+1) \chi_j(\mu_{\text{\tiny{SO(3)}}}) e^{\beta \mu \frac{Q}{e}} \int Dg_{\mu \nu} D\chi \hspace{0.1cm}e^{- I_{Q,j}[g_{\mu \nu},\chi]},
\eeq
where $\chi_j(\theta)={\small \frac{\sin (2j+1)\theta}{\sin \theta}} $ is the $SO(3)$ character. The sum is over integer spin since we chose the group to be $SO(3)$, and would be over half-integer values had we chosen $SU(2)$. The gravitational action includes extra terms in the dilaton potential from the integrated out gauge fields 
\bea
\label{eq:action-Q,J-sector}
I_{Q,j}[g,\chi]&=& -\frac{1}{4G_N} \int_{M_2}d^2 x \sqrt{g} \left[\chi R -2  U_{Q,j}(\chi)\right] -\frac{1}{2 G_N} \int_{\partial M_2} du \sqrt{h} \chi K , \\
U_{Q,j}(\chi) &=&r_0\left[ \frac{G_N }{4 \pi  \chi^{3/2}}  Q^2+ \frac{3 G_N^2}{ \chi^{5/2}}j(j+1) - \frac{3 \chi^{1/2}}{L^2} - \frac{1}{\chi^{1/2}}   \right]\,.
\ea
Fixing the field strength (which corresponds to studying the system in the canonical ensemble) instead of the gauge field holonomy (the grand canonical ensemble) simply isolates individual terms in the sum over $Q$ and $j$ which corresponds to fixing the black hole charge and, as we will show shortly, to its angular momentum. 

The equations of motion corresponding to this theory are given by \cite{Grumiller:2007ju}
\bea
\nabla_\mu \nabla_\nu \chi - g_{\mu\nu} \nabla^2 \chi - g_{\mu\nu} U_{Q,j}(\chi)&=&0\\
R - 2 \partial_\chi U_{Q,j}(\chi) &=& 0.
\ea
By fixing part of the gauge freedom, the most general static solution can be put into the following form 
\beq\label{dilgravansatz}
\chi=\chi(r),~~~~ds^2 =\frac{\chi^{1/2}}{r_0} \left[ f(r) d\tau^2 + \frac{dr^2}{f(r)}\right].
\eeq
The equation for the dilaton gives $\partial_{r^2} \chi = {\rm constant}$, and using remaining gauge freedom the solution can be put in the form $\chi(r) =r^2$. For this choice the metric equation becomes 
\beq
f(\chi) =\frac{1}{\chi^{1/2}}\left[ C - \frac{1}{2r_0} \int^\chi d\chi U_{Q,j}(\chi)\right],
\eeq
where $C$ is an integration constant that can be fixed by the boundary conditions. This gives the complete solution of the dilaton gravity equations. After analyzing some particular cases, we will see why the specific ansatz \eqref{dilgravansatz} that we chose is convenient. 

First, the simplest case is to study states with $j=0$. Then the equation of motion for the metric and dilaton for each effective action \eqref{eq:action-Q,J-sector} yields 
\beq
f(\chi) =\frac{1}{\chi^{1/2}}\left[ C - \frac{1}{2r_0}\int^\chi d\chi U_{Q,0}(\chi)\right] = 1+ \frac{\chi}{L^2} + \frac{G_N}{4\pi }\frac{Q^2}{\chi} + \frac{C}{\chi^{1/2}} .
\eeq
Using $\chi=r^2$ and the boundary conditions at large $r$ we can fix the integration constant $C=-2 G_N M$. Replacing this in the equation above, and replacing the two dimensional metric \eqref{dilgravansatz} into the four dimensional \eqref{eq:beyond-S-wave-metric-ansatz}, we see that this precisely agrees with the Reinsner-Nordstr\"{o}m solution \eqref{eq:RN-black-hole-AdS} described in section \ref{sec:setup} for fixed charge $Q$.  

We can now discuss the case of arbitrary small $j$. Up to subtleties about the backrection of the $SO(3)$ gauge field on the $g_{rr}$ and $g_{\tau \tau}$ metric components, the states with fixed $j$ can be identified as the KN solutions reviewed in appendix \ref{app:SO(3)-gauge-field-review}. Specifically, as we show in appendix \ref{app:SO(3)-gauge-field-review}, the deformation from Reissner-Nordstr\"{o}m \eqref{eq:RN-black-hole-AdS} is given by  $SO(3)$ gauge field solutions, plugged into the metric ansatz \eqref{eq:beyond-S-wave-metric-ansatz}: 
\be 
\label{eq:deformation-from-RN-to-KN}
g_{\mu \nu} = g_{\mu \nu}^{\rm RN} + \delta g_{\mu \nu} \,, \qquad 
\delta g_{\mu \nu} dx^\mu dx^\nu = 4i  r^2 \sin^2 \theta \Big(\a_1 + \frac{\a_2}{r^3} \Big) d\phi d\tau\,.
\ee 
$\a_1$ and $\a_2$ are two constants which are determining by the boundary conditions on the $SO(3)$ gauge field and by requiring that the gauge field be smooth at the black hole horizon.  Turning on a non-trivial profile for the $SO(3)$ gauge field as in \eqref{eq:deformation-from-RN-to-KN} breaks the $SO(3)$ rotational isometry  down to $U(1)$. This is the same as in the well-known KN solution reviewed in appendix \ref{app:SO(3)-gauge-field-review}. Solving the equations of motion in the semiclassical limit when fixing the field strength on the boundary to $H^3_{r\tau} |_{\partial M_2} =i \frac{6 G_N j^2}{\sqrt{2} r^4}|_{\partial M_2} $, corresponds to fixing $j$ in the sum in \eqref{eq:part-function-sum-over-Q,J}, and yields a solution with a fixed 4d total angular momentum $J = j$.\footnote{Where $J$ is normalized as in the KN solution \eqref{metext-KN-BHs-AdS}. } Since the KN solution is the unique solution with a $U(1)$ rotation isometry and with fixed angular moment and charge, this makes the metric ansatz that includes the deformation \eqref{eq:deformation-from-RN-to-KN} agree (for sufficiently small $j$) with the KN solution up to diffeomorphisms. 

We can now address the subtlety about the $SO(3)$ gauge field backreacting on the $g_{rr}$ and $g_{\tau \tau}$ components of the metric. The reason why we need to account for such backreaction is that it can source other massive Kaluza-Klein modes of the metric, which are not accounted for in the action \eqref{eq:action-Q,J-sector}. In order to understand the $SO(3)$ gauge field backreaction, we can repeat the analysis above in which we studied the backreaction of the $U(1)$ gauge field on $f(r)$. For $j\neq 0$ we get a correction to the metric $\delta_j f \sim \frac{G_N^2 j(j+1)}{r^4}$. Since we do not want to source further backreaction on the massive Kaluza-Klein modes, we will require that this correction is small everywhere far from the horizon and require that the spin of the black hole satisfy $j(j+1) \ll (r_h/\ell_{Pl})^{4}$.

\subsection{New boundary conditions in the throat}
\label{sec:bdy-conditions-in-the-throat}
 
 While quantizing the action \eqref{eq:action-Q,J-sector} directly is out of reach, we can do better by separating the integral in the action in the NHR and FAR. To conveniently manipulate the action into a form where quantization can be addressed, we follow the strategy of \cite{Nayak:2018qej}. Namely we choose the NHR and FAR to be separated by an arbitrary curve with a fixed dilaton value $\chi|_{\partial M_{\rm NHR}} = \chi_b$ and fixed intrinsic boundary metric $h_{uu} = {1}/{\e^2}$ and proper length $\ell = \int du \sqrt{h}$.  

In the NHR, the equations of motion fixes the value of the dilaton at the horizon to be 
\be 
\Phi_0 \equiv \frac{\chi(r_h)}{G_N} = \frac{r_0^2(Q)}{G_N} \,,
\ee
which acts as a very large constant background. The function $r_0(Q)$ obtained from dilaton-gravity is equivalent to solving \eqref{extpar}. In  the  NHR where $r-r_h \ll r_h$ we can study small fluctuations around this value $\chi(r) /G_N= \Phi_0 + \Phi(r)$. Expanding the action to first order we find that 
\be
I_\text{NHR}^{Q, j}[g_{\mu \nu}, \chi] &=  \frac{1}{4}\int_{M_\text{NHR}} d^2 x \sqrt{g} \left[-\Phi_0 R  - \Phi \Big(R + \frac{2}{L_2^2}\Big) + O\left(\frac{\Phi^2}{\Phi_0}\right)\right] \,,
\ee 
where the two dimensional AdS radius is $L_2=\frac{Lr_0}{\sqrt{L^2+6 r_0^2}}$, which in general (except for the case of large black holes in AdS$_4$) also depends on the charge of the black hole through $r_0(Q)$. From now, $L_2$ and $r_0$ should be understood as functions of the charge. The last term captures a quadratic correction in the dilaton variation. The quantization of the above action has been widely discussed in the presence of an appropriate boundary term. 

\begin{figure}[t!]
\begin{center}
\begin{tikzpicture}
\draw [purple,  thick]  (0,-1.0) -- (0,1.0);
\node[text width=2cm] at (1.2,0) {\small Horizon};
 \draw[red, thick] (0, -1.0) .. controls (5.5,-1.05) .. (6, -1.35);
   \draw[red, thick] (6, -1.35) .. controls (7,-1.5) .. (10, -3.5);
 \node[text width=2cm] at (3,2.2) {\begin{center}{ \textit{NHR} \\ AdS$_2 \times S^2$ \\
 JT gravity }\end{center}};
  \node[text width=2cm] at (9.95, 2.0) {\begin{center}{\textit{FAR}  \\ Extremal black hole saddle}\end{center} };
  \node[text width=3cm] at (7.65, 0) {\begin{center}{\small Boundary ($\partial \cM$) \\ theory: \\ GHY term}\end{center}};
    \node[text width=3cm] at (5.8, -1.55) {\begin{center}{$\delta K \sim K_{\rm NHR} - K_{\text{ext}}$}\end{center} };
   \draw[red, thick] (0, 1.0) .. controls (5.5,1.05) .. (6, 1.35);
  \draw[red, thick] (6, 1.35) .. controls (7,1.5) .. (10, 3.5);
   \draw[blue,thick] plot [smooth,tension=0.6] coordinates {(6,-1.35)(6.2,-0.5) (6.15,-0.25) (6, 0)(5.8,0.5)(6,1.35)};
    \draw[ thick, ->] {(6,1.35)  --  node[above] {$\hat n_\text{NHR}\qquad$}(6.8,2.0)};
        \draw[ thick, ->] {(6,1.35)  --  node[above] {$\hat n_\text{FAR}$}(5.,1.25)};
\end{tikzpicture}
\end{center}
\caption{A cartoon of the near-horizon region (NHR) and the far-away region (FAR) separated by a curve along which the boundary term of JT gravity will need to be evaluated. }\label{fig:regions-disc}
\end{figure}
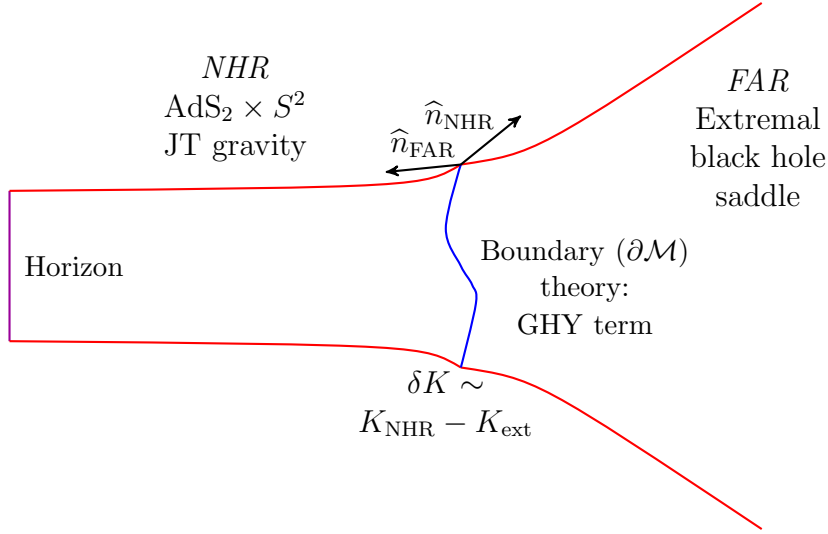

We will see next how this boundary term arises from including fluctuations in the FAR region. We proceed by expanding the near-extremal metric and dilaton in the FAR region into their contribution from the extremal metric and their fluctuation:
\be
\label{eq:expansion-in-far-away-region}
g_{\mu \nu} &= g_{\mu \nu}^\text{ext} + \delta g_{\mu \nu}^\text{near-ext}\,, \qquad &\chi = \chi^{\extr} + \delta \chi^{\nextr} \,.
\ee
Both the extremal and near-extremal $4d$ metrics are solutions to the equations of motion at fixed $\b$, i.e. with periodic Euclidean time $\tau \sim \tau+\b$. The extremal solution however contains a singularity at the horizon if imposing any periodicity for the Euclidean time. Nevertheless, if separating the space into the NHR and the FAR, the singularity would not be present in the latter region and we can safely expand the action around the extremal solution.  If expanding around the the extremal metic, following from the variational principle the first order term in the expansion is solely a total derivative term which when integrated by parts results in a total boundary term.  Explicitly, the action is given by  
\be 
&I_{\rm FAR}^{Q, j}[g_{\mu \nu} , \chi]  = I_{\rm FAR}^{Q, j}[g_{\mu \nu}^\text{ext}, \chi^\text{ext}] - \frac{1}{2 G_N}\int_{\partial M_{\rm NHR}} du \sqrt{h} \left[ \chi\delta K - (\partial_n \chi - \chi K) \delta \sqrt{h_{uu}}  \right] \,, \nn \\  &\delta K \equiv K_{\rm NHR} - K_{\rm ext}\,, \qquad \delta \sqrt{h_{uu}} = 0\,.
\ee
The last equality follows from the fact that we have imposed Dirichlet boundary conditions for the intrinsic boundary metric. Consequently, as sketched in figure \ref{fig:regions-disc}, we obtained a surface which has a small discontinuity precisely on the curve that separates  the NHR from the FAR. Above, $K_{\rm NHR} $ is the extrinsic curvature evaluated on the boundary of the NHR (defined with respect to the direction of the normal vector $\hat n_{\rm NHR}$) and $K_\text{ext}$ is the extrinsic curvature evaluated on the boundary of the FAR with the extremal metric on it (wrt the normal vector $\hat n_{\rm FAR}$).

We can now understand the effect of the Dirichlet boundary conditions for the dilaton $\chi_b =G_N( \Phi_0+\Phi_b/(2\e))$ and proper boundary length $ \ell = \int du \sqrt{h} = \b L_2/\e$. Here, $\e$ is some parameter fixed by the value of  $\ell$ and $\b$ whose role we will understand shortly. Curves of constant  dilaton in the extremal solution are fixed to have a constant value of $ r_{\partial M_{\rm NHR}} \equiv r_0 + \delta r_\bdy$ and are parametrized by $\tau$ when using the coordinate system in \eqref{eq:extremal-metric}. In the extremal solution, the dilaton value, proper length and extrinsic curvature $K_{\rm  ext}$ on the extremal side are all fixed by the value of $\delta r_\bdy$:
\be
\label{eq:AdS2-scales-and-bdy-cond}
&\chi_b  = G_N\Big( \Phi_0 + \frac{\Phi_{b, Q}}{2\e} \Big)\,, \qquad \text{with} \qquad \frac{\Phi_{b, Q}}{2\e} =  \frac{r_0 \delta r_\bdy}{G_N}  \,,\nn \\  &\ell = \int du \sqrt{h} = \frac{\b L_2}{\e}\,,  \qquad \text{with} \qquad  \qquad \e  = \frac{L_2^2}{ \delta r_\bdy}\,, \qquad\Phi_{b,Q} =  M_{{\text{\tiny{SL(2)}}}}^{-1} = \frac{r_0L_2^2}{G_N}  \,,\nn \\ & K_\text{ext} = \frac{1}{L_2} \Big(1 - \frac{4}3 \frac{\delta r_\text{bdy}}{r_0} + \frac{(L^2 + 25 \delta r_\text{bdy}^2)}{(12 r_0^2)} + O\left(\frac{\delta r_\text{bdy}^3}{r_0^3}\right) \Big)\,,
 \ee
where we computed the extremal extrinsic curvature using the metric \eqref{eq:extremal-metric}. In the near-extremal limit we have that $\b \gg\e$ and $\Phi_b \gg \e$. These inequalities will prove important in relating \eqref{eq:AdS2-scales-and-bdy-cond} to a boundary Schwarzian theory. 

 We see here explicitly that the renormalized value of the dilaton is precisely given by the inverse mass gap scale in the way defined previously by thermodynamic arguments. Consequently, the overall action is given by 
\be
\label{eq:simplified-action}
I_{EM}^{Q, j} =  &-\frac{1}{4}\int_{M_\text{NHR}} d^2 x \sqrt{g} \left[\Phi_0 R  + \Phi \Big(R + \frac{2}{L_2^2}\Big) \right] \nn \\ &-\frac{1}2 \int_{\partial M_{\rm NHR}} du \sqrt{h} (\Phi_0 + \Phi) \Big[K_{\rm NHR}  -\frac{1}{L_2}\Big(1 + \frac{4}3 \frac{\delta r_\text{bdy}}{r_0}\Big)\Big]  + I_{\rm FAR}^{Q, j}[g_{\mu \nu}^\text{ext}, \chi^\text{ext}]\,.
\ee 

The quadratic fluctuations in the FAR region are suppressed compared to the contribution of the first two NHR terms in  \eqref{eq:simplified-action}.\footnote{Even when we will integrate over order one fluctuations of the Schwarzian mode in the next section, the fluctuations in the metric near the boundary of AdS$_2$ is suppressed by the cut-off. For example $\delta g_{\tau\tau} \sim \epsilon^2 {\rm Sch}(\tau,u)$. Therefore fluctuations in the FAR region are always small, and become large only very close to the horizon far inside the throat.} Therefore, we will neglect the possible quadratic (or higher order) fluctuations around the extremal metric in the FAR region and proceed by evaluating the contribution of FAR action on-shell. To simplify the computation, we will, for now, focus on the $j=0$ sector where there is no backreation from the $SO(3)$ gauge field on the other components of the metric. On-shell, the bulk term in the FAR action evaluates to
\be 
\label{eq:on-shell-FAR-action-evaluates-to}
I_{\rm FAR, \text{ bulk}}^{Q, j = 0}[g_{\mu \nu}^\text{ext}, \chi^\text{ext}] &=- \frac{1}{4 G_N} \int d^2x \sqrt{g_\text{ext}} \left[\chi R - 2 U_{Q, 0}[g_\text{ext}^{\mu \nu}, \chi^{ext}]\right] \nn \\ & = -\frac{3 r_{\partial M_2} \b}{4 G_N} \Big(1 + \frac{r_{\partial M_2}^2}{12 L^2} \Big) + \frac{2r_0 \b}{G_N} \Big(1+ \frac{2r_0^2}{L^2}\Big) - \frac{\b \delta r_\text{bdy}}{2G_N}\Big(1+\frac{6r_0^2}{L^2}\Big)    \,.
\ee
where, as we will see shortly, the divergent terms can be canceled by adding counter-terms to the boundary term in the action \eqref{eq:Einstein-Maxwell-action} (which we have so far neglected). We now include this boundary term from \eqref{eq:Einstein-Maxwell-action} (associated to the Dirichlet boundary conditions on $\partial M_2$) together with possible counter-terms. This evaluates to:
\be 
\label{eq:boundary-term-evaluation}
I_{\rm FAR, \text{ bdy.}}^{Q, j = 0} &[g_{\mu \nu}^\text{ext}, \chi^\text{ext}]=\frac{1}{2 G_N} \int_{\partial M_2} du \sqrt{h} \Big( \chi K + \cC_1 \frac{\chi^{3/4}}{r_0^{3/2}}+ \cC_2 \frac{r_0^{1/2}}{\chi^{1/4}}\Big) \nn \\ &=  \frac{\beta  r_{\partial M_2}^3 \left(2 \cC_1 L-3 r_0^2\right)}{4 G_N L^2
   r_0^2}+\frac{\beta  r_{\partial M_2}  \left(\cC_1 L^2+2 \cC_2r_0^2-L
   r_0^2\right)}{4 G_N L r_0^2}-\frac{\beta  \cC_1 \left(L^2+2
r_0^2\right)}{2 G_N L r_0} \,, 
\ee
where the terms including $\cC_1$ and $\cC_2$ are the counterterms necessary to cancel the divergence in \eqref{eq:on-shell-FAR-action-evaluates-to}.  In order to cancel the divergence in \eqref{eq:on-shell-FAR-action-evaluates-to} we set, 
\be
\cC_1 =\frac{2r_0^2}{L}  \,, \qquad \cC_2 = L \,. 
\ee
We can also find precisely the same terms with the right prefactors by dimensionally reducing the holographic counterterm of \cite{Emparan:1999pm}, reproducing the same overall on-shell action. In total we thus find that 
\be
\label{eq:FAR-action-almost-total}
I_{\rm FAR}^{Q, j = 0} &= I_{\rm FAR, \text{ bulk}}^{Q, j = 0} [g_{\mu \nu}^\text{ext}, \chi^\text{ext}] + I_{\rm FAR, \text{ bdy.}}^{Q, j = 0} [g_{\mu \nu}^\text{ext}, \chi^\text{ext}] =\frac{r_0 \b}{G_N} \Big(1+ \frac{2r_0^2}{L^2}\Big) - \frac{\b \delta r_\text{bdy}}{2G_N}\Big(1+\frac{6r_0^2}{L^2}\Big)  \nn \\ & = \b M_0(Q) - \frac{\b \delta r_\text{bdy}}{2G_N}\Big(1+\frac{6r_0^2(Q)}{L^2}\Big) \,,
\ee
where in the last line we emphasize the charge dependence of the extremal mass and horizon radius, given by \eqref{eq:extremalparam}.  The $\delta r_\text{bdy}$ dependent term in the action \eqref{eq:FAR-action-almost-total}, $\frac{2\sqrt{6}}{3 G_N}\int_{\partial M_{\rm NHR}} du \sqrt{h} \frac{\chi \delta r_\bdy}{r_0}  $,  also precisely cancels the $\delta r_\text{bdy}$ term in \eqref{eq:on-shell-FAR-action-evaluates-to}. This is simply a consequence of the fact that the parameter $\delta r_\bdy$ is chosen arbitrarily to separate $M_2$ into the NHR and the FAR and, consequently, the fact that all our results are independent of $\delta r_\bdy$ can be seen as a consistency check. 

Next, we can consider the contribution to the action of the $SO(3)$ gauge fields and of the backreaction of the field on other components of the metric. Corrections could appear in the contribution to the partition function in the extremal area term or in the extremal energy. The former is of order $\delta \Phi_0 \sim \frac{G_NL^2}{r_0^4}j(j+1)$ (for a large black hole in AdS) or $\delta \Phi_0 \sim \frac{G_N}{r_0^2}j(j+1)$ (for a black hole in flat space) and therefore is very small and can be neglected in either case. The term coming from the correction to the extremal mass, originating from the backreaction on the metric and by the $SO(3)$ Yang-Mills term in the action, is multiplied by a large factor of $\beta$ and gives the leading correction
\beq
\label{eq:correction-from-SO3-gauge-field}
M_0(Q, j)  =  M_0(Q, j=0) + \frac{G_N}{2r_0^3} j(j+1) +O(j^4),
\eeq
where $r_0(Q)$ is the extremal horizon size for the RN black hole given by \eqref{extpar} \footnote{\textbf{v2:} For a more accurate account of this correction, see section 4.4 of \cite{HITZ}.}. In principle, the backreaction of the $SO(3)$ gauge field also affects the boundary value of the dilaton $\Phi_b/(2\epsilon)$. However, such a contribution appears at the same order as other $O(1/\Phi_0)$ corrections, which we have ignored in the NHR. Therefore, we will solely track the $Q$-dependence of $\Phi_b(Q,j) \to \Phi_{b, Q}$.  

 
 Thus, in total we find that the dynamics of the near-extremal black hole is described by
\be 
\label{eq:simplified-effective-action-final}
I_{EM}^{Q, j}[g_{\mu \nu}^\text{ext}, \chi^\text{ext}]  & =   \b M_0(Q, j) -\frac{1}{4}\int_{M_\text{NHR}} d^2 x \sqrt{g} \left[\Phi_0(Q, j) R  + \Phi \Big(R + \frac{2}{L_2^2}\Big) + O\left(\frac{\Phi^2}{\Phi_0^2}\right) \right] \nn \\ &-\frac{1}2 \int_{\partial M_{\rm NHR}} du \sqrt{h} \left[\Phi_0(Q, j) \,K_{\rm NHR}+ \frac{\Phi_{b, Q}}{\e}\Big(K_{\rm NHR}  -\frac{1}{L_2}\Big) \right] \,,
\ee
where the on-shell contribution of the FAR action can be seen as an overall shift of the ground state energy of the system. We can now proceed by using \eqref{eq:on-shell-FAR-action-evaluates-to} to determine the exact ground state energy of the system, and then by quantizing the remaining degrees of freedom in \eqref{eq:simplified-action}.

Before moving on, we can briefly comment on corrections coming from non-linearities in the dilaton potential present in the first line of \eqref{eq:simplified-effective-action-final}. To leading order, we get the JT gravity action written above. The next correction behaves like $\delta U \sim  \Phi^2/\Phi_0$. The contribution to the partition function from such a term was computed in \cite{Kitaev:2017awl} and scales as $\delta \log Z \sim \Phi_b^2/(\beta^2 \Phi_0)$. Such a contribution is suppressed by the large extremal area $\Phi_0 \gg 1$. Higher-order corrections to the dilaton potential are further suppressed by higher powers of $\Phi_0$ and, more importantly, decay faster at low temperatures. Therefore, they can all be neglected.

\section{The partition function for near-extremal black holes}
\label{sec:part-function-large-BHs-AdS4}
\subsection{An equivalent 1D boundary theory}
\label{sec:equivalent-boudnary-theory}

We will now evaluate the contribution to the partition function of the quantum fluctuations from the remaining graviton and dilaton fields present in the effective action of the NHR \eqref{eq:simplified-effective-action-final}. We briefly review this procedure by first reducing the path integral of \eqref{eq:simplified-effective-action-final} to that of a boundary Schwarzian theory. 

Integrating out the dilaton enforces that the curvature is fixed to $R = -2/L_2^2$.\footnote{In order to enforce such a condition, the contour for dilaton fluctuation $\Phi(x)$ needs to go along the imaginary axis such that 
\be 
\int Dg_{\mu \nu} \int_{\Phi_b - i \infty}^{\Phi_b + i \infty } D\Phi ~e^{\int_{M_{NHR}} d^2 x \sqrt{g} \Phi\big(R -\frac{2}{L_2^2}\big)}  =  \int Dg_{\mu \nu} \delta\Big(R - \frac{2}{L_2^2}\Big)\,.
\ee
This choice of contour for $\Phi$ isolates the same type of constant curvature configurations in Euclidean signature as those that dominate in the Lorentzian path integral. More details about this choice of countour in the context of near-extremal black holes are discussed in footnote 9 of \cite{Iliesiu:2019lfc}.} Thus, each near-horizon region configuration that contributes to the path integral is a patch of $AdS_2$ cut along a curve with a fixed proper length $\ell$.  Following \cite{Maldacena:2016upp}, we can write the $AdS_2$ metric by $ds_{AdS_2}^2  =  L_2^2 \frac{dF^2 + dz^2 }{z^2}$ and parametrize the boundary with a proper time $u$, with $u \in [0, \b)$ and $h_{uu} = 1/\e^2$. In this case, one can solve for the value of $z(u)$ in terms of $F(u)$ on the boundary, in the limit in which $\b \gg \e$ to find that $z(u) = \e F'(u)$. The extrinsic curvature can then be written in terms of the Schwarzian derivative  \cite{Maldacena:2016upp}:
\be
K_\text{NHR}  = \frac{1}{L_2} \left[1+ \e^2\, \Sch(F, u) + O(\e^4)\right]\,,\qquad \Sch(F, u) = \frac{F'''}{F'} - \frac{3}2 \left(\frac{F''}{F'}\right)^2 \,.
\ee
The geometry we are working with in the NHR after reducing on $S^2$ is actually the hyperbolic disk. We can easily go from the Poincare coordinates to the disk by replacing 
\beq
F(u) = \tan \frac{\pi \tau(u)}{\beta},~~~~\tau(u+\b) = \tau(u) + \b 
\eeq
in the Schwarzian action. Here $\tau$ parametrizes the Euclidean circle at the boundary of the NHR which we glue to the FAR region. For simplicity we will mostly write the Schwarzian action in terms of $F(u)$ instead. 

The path integral over the the metric reduces to an integral over the field $F(u)$ and the partition function becomes:\footnote{Above, the path integral measure $\cD\mu[F]$ over the field $F(u)$ can be determined from the symplectic form associated to an $SL(2, \mR)$ BF-theory which is equivalent on-shell to JT gravity. } 
\be 
\label{eq:partition-function-sum-over-q-and-j}
Z_{\rm RN}[\b, \mu, \mu_{\text{\tiny{SO(3)}}}] &= \sum_{Q \in e\cdot \mZ, j \in \mZ} (2j+1) \chi_j( \mu_{\text{\tiny{SO(3)}}}) e^{-\frac{Q}{e}\beta \mu} e^{\pi \Phi_0(Q, j)} e^{-\b M_0(Q, j)} \nn \\ &\qquad \qquad \times \int \frac{\mathcal{D}\mu[F]}{{\rm SL}(2,\mathbb{R})} e^{\Phi_{b, Q} \int_0^\beta  du ~\Sch(F, u)}\,.
\ee 
This relation shows that we can identify the term giving the extremal area $S_0 = \pi \Phi_0$ coming from the topological part of the dilaton gravity NHR action. The extremal mass term comes from the action in the FAR region. The path integral over the Schwarzian theory includes finite temperature near-extremal effects. The effective coupling of this mode depends on the charge and spin of each black hole in the ensemble. 

Before reviewing the quantization of \eqref{eq:partition-function-sum-over-q-and-j}, it is also interesting to study the possibility that the sum over all the possible representations is reproduced by a single 1d theory.
 Reproducing the sum over charges can be done by coupling the Schwarzian theory to a theory having a $U(1)\times SO(3)$ symmetry. As explained in \citep{Anninos:2017cnw, Iliesiu:2019lfc}, the theory that exhibits this symmetry and correctly captures the sum over charges is that of a particle moving on a $U(1) \times SO(3)$ group manifold. To obtain this model, we  introduce four additional fields: a compact scalar $\theta(u) \sim \theta(u) +2\pi$ together with a Lagrange muliplier $\a(u)$ and a field $h(u) \in  SO(3)$ together with another Lagrange multiplier $\pmb \ma(u) \in SO(3)$. The general coupling between the particle moving on a group manifold and the Schwarzian theory is given by:
\be 
\label{eq:particle-moving-on-group-manifold-general}
I_{\text{Sch} \times U(1) \times SO(3) } = -\int_0^\b du \Big[i  \alpha \theta' + i \tr \left(\pmb \ma h^{-1} h'\right) + \cV(\a, \tr \pmb \a^2)- \cW(\a) \,\Sch\left(F, u\right) \Big]\,,
\ee
where the potential  $\cW(\a) $ is independent of the the $SO(3)$ degrees of freedom since we are neglecting the effect of angular momentum of the boundary value of the dilaton $\Phi_{b, Q}$.
 
When the generic potential $\cV(\a,  \tr \pmb \a^2)$ is of trace-class, the theory has a $U(1)$ symmetry $\theta \to \theta + a$ and two $SO(3)$-symmetries generated by the transformations $h \to g_L h g_R$ and $\pmb \a \to g_R^{-1} \pmb  \a g_R $, with $g_L, \, g_R \in SO(3)$.  Consequently, the Hilbert space arranges itself in representations of $U(1) \times SO(3) \times SO(3)$. However, the quadratic Casimir of both $SO(3)$-symmetries is in fact the same. Therefore, the Hilbert space arranges itself in  representations of $U(1)$ and two copies of the same $SO(3)$-representation. If we are interested in reproducing the near-extremal black hole partition function with Dirichlet boundary conditions for the $U(1)$ and $SO(3)$ gauge fields, then we need to introduce a chemical potential for the $U(1)$ symmetry of \eqref{eq:particle-moving-on-group-manifold-general} and for one of its $SO(3)$ symmetries.  This can be done by introducing a $U(1)$ background gauge field, $\cA$ with $\exp (\oint \cA) = e^{\beta \mu}$, and an $SO(3)$ background gauge field, $\cB$ with $\cP\exp(\oint \cB) \sim e^{i \beta \mu_{\text{\tiny{SO(3)}}} \sigma_3}$, coupling the first background to the $U(1)$ charge through $-i \int_0^\b du \a \cA_u $ and the second background to the $SO(3)$ charges through $-i\int_0^\b du\, \Tr(\pmb \a \cB_u)$.  In such a case, the partition function of the general theory \eqref{eq:particle-moving-on-group-manifold-general} can be shown to be \cite{Iliesiu:2019lfc}:\footnote{When taking the trace over the Hilbert space of the theory \eqref{eq:particle-moving-on-group-manifold-general}  and summing over states within the two copies of some $SO(3)$ representation $j$ then the sum over the gauged copy yields $\chi_R(\theta)$ while the sum over the other copy yields the degeneracy $\dim R$ in \eqref{eq:general-partition-function}. }
\be 
\label{eq:general-partition-function}
Z_{Sch\times U(1) \times SO(3)} =\hspace{-0.3cm} \sum_{Q \in e \cdot \mZ, j \in \mZ} e^{\beta \mu \frac{Q}{e}} (2j+1)\chi_j( \mu_{\text{\tiny{SO(3)}}}) e^{-\beta \cV(\frac{Q}{e}, j(j+1))} e^{\cW(\frac{Q}{e}, j(j+1))\int_0^\b du\hspace{0.1cm} \Sch(F, u)}\,,
\ee
which up to an overall proportionality constant corresponding to the extremal black hole entropy  agrees with the form of \eqref{eq:partition-function-sum-over-q-and-j}. Therefore, the potentials $\cV(\a, \pmb \a)$ and $\cW(\a, \pmb \a)$ need to be tuned in order for the partition function of the theory \eqref{eq:particle-moving-on-group-manifold-general} to reproduce the charge dependence in the sum in \eqref{eq:partition-function-sum-over-q-and-j}. For example, for large black holes in $AdS_4$ we find that:
\be 
\cV(\a, \pmb \a) =
\frac{|\a|^{3/2}}{(3\pi)^{3/4}(2L)^{1/2} G_N^{1/4} }   +\frac{\sqrt{2}G_N^{1/4}(3\pi)^{3/4}}{L^{3/2} |\a|^{3/2}} \tr \pmb \a^2 \,, \qquad \cW(\a)= \frac{|\a|^{1/2} L^{5/2}}{6 \sqrt{2}(3\pi G_N^{3})^{1/4} }\,.
\ee 
For black holes in flat space we find: 
\be 
\cV(\a, \pmb \a) =\frac{|\a|}{2(\pi  G_N)^{1/2}} +\frac{4\pi^{3/2}}{G_N^{1/2}|\a|^{3}}  \tr \pmb \a^2  \,, \qquad \cW(\a)= \frac{|\a|^3 G_N^{1/2} }{8 \pi^{3/2} }\,.
\ee 
We will see in the next section that for fluctuations around extremality the action for the U(1) and SO(3) mode further simplifies.

\subsection{The partition function at $j=0$}
\label{sec:large-BHs-AdS4-canonical-ensemble}
We have identified the effects that dominate the temperature dependence in the near-extremal limit. In this section, we will put everything together to find a final answer for the partition function. To at first simplify the discussion, we will pick boundary conditions in the four-dimensional theory that fix the angular momentum $j$ to zero. In the dimensional reduced theory this is equivalent to picking only the $j=0$ sector of expression \eqref{eq:partition-function-sum-over-q-and-j}. We will analyze fixed $U(1)$ charge and chemical potential separately. 

\begin{center}
\textbf{Fixed Charge}
\end{center}
This is the simplest case to consider where we fix the temperature, $U(1)$ charge $Q$ and angular momentum to zero. From a Laplace transform of equation \eqref{eq:partition-function-sum-over-q-and-j} the partition function is given by 
\be 
\label{eq:partition-function-fixed-q}
Z_{\rm RN}[\b, Q] = e^{\pi \Phi_0(Q)} e^{-\b M_0(Q)}  \int \frac{\mathcal{D}\mu[F]}{{\rm SL}(2,\mathbb{R})} e^{\Phi_{b, Q} \int_0^\beta  du \hspace{0.1cm}\Sch(F, u)}.
\ee 
This means that for boundary conditions of fixed charge, the $U(1)$ mode is effectively frozen and does not contribute to the partition function, leaving only the Schwarzian mode. The path integral of the Schwarzian theory can be computed exactly and gives 
\beq\label{eq:schwarzianpf}
Z_{\rm Sch}(\Phi_{b,Q}, \beta) \equiv  \int \frac{\mathcal{D}\mu[F]}{{\rm SL}(2,\mathbb{R})} e^{\Phi_{b, Q} \int_0^\beta  du \hspace{0.1cm}\Sch(F, u)}= \Big( \frac{\Phi_{b,Q}}{\beta}\Big)^{3/2} e^{\frac{2\pi^2 }{\beta}\Phi_{b,Q_0}}.
\eeq
Then the final expression for the canonical partition function is 
\be 
\label{eq:RN-partition-function-canonical-ensemble}
Z_{\rm RN}[\b, Q] = \Big( \frac{\Phi_{b,Q}}{\beta}\Big)^{3/2} e^{\pi \Phi_0(Q)-\b M_0(Q)+\frac{2\pi^2 }{\beta}\Phi_{b,Q_0}}.
\ee 
Here the first term comes from the gravitational one-loop correction from the JT mode which dominates at low temperatures. This gives a correction $-\frac{3}{2}T\,\log T$ to the free energy (equivalently a $\frac{3}{2} \log T$ correction to $\log{Z}$). The terms in the exponential are first the extremal entropy through $S_0 = \pi \Phi_0$, the extremal mass term $-\beta M_0(Q)$ and the third gives the leading semiclassical correction near extremality. The temperature dependence of this expression is exact even for $\Phi_{b,Q}/\beta$ finite. The result is valid as long as, stringy effects are not important, $r_0 \gg \lpl$ (equivalently, $Q \gg 1$) and when the black hole is near-extremal, $\b \gg r_0$ $\left(\text{equivalently, }\b^2 \gg \frac{L^2}{6}\Big[\sqrt{1+\frac{3G_N Q^2}{\pi L^2}}-1\Big]\right)$. 

With this expression we can analyze the thermodynamics of the system. The entropy is given by 
\bea
S(\beta,Q) &=& (1- \beta \partial_\beta) \log Z = S_0 + \frac{4 \pi^2 \Phi_{b,Q}}{\beta} - \frac{3}{2}\log \frac{\beta}{e \Phi_{b,Q}},\\
E(\beta,Q) &=& M_0 + \frac{2 \pi^2 \Phi_{b,Q}}{\beta^2} + \frac{3}{2\beta}
\ea
This gives a resolution of the ``thermodynamic gap scale" puzzle. At very low temperatures the energy goes as $E-M_0 \sim \frac{3}{2} T$ (as opposed to $\sim T^2$). Therefore the energy is always bigger than the temperature and the argument of \cite{Preskill:1991tb} does not apply. We will see this again in the next section when we work directly with the density of states, showing explicitly that there is no gap in the spectrum.

Finally, there are well-known corrections to the partition function of an extremal black hole computed by Sen \cite{Sen:2011ba} coming from integrating out matter fields. Those effects can correct the extremal entropy $S_0$ at subleading orders. These corrections are significant compared to the ones coming from the Schwarzian mode but are temperature-independent in the limit we are taking (see also the results of \cite{Charles:2019tiu}) and can be absorbed by a shift of $S_0$. As previously stated, the goal of this paper is to study the leading temperature-dependent contributions to the free energy. Therefore, we can neglect these possible shifts of $S_0$.   

\begin{center}
\begin{minipage}{\textwidth}
\begin{center}
\textbf{Fixed Chemical Potential}
\end{center}~
The partition function with fixed $U(1)$ chemical potential $\mu$ and zero angular momentum is given by 
\be 
\label{eq:partition-function-sum-over-q}
Z_{\rm RN}[\b, \mu] = \sum_{Q \in e \cdot  \mZ} e^{\beta \mu \frac{Q}{e}} e^{\pi \Phi_0(Q)} e^{-\b M_0(Q)}  Z_{\rm Sch}(\Phi_{b,Q}, \beta) 
\ee 
\end{minipage}
\end{center}
As previously mentioned the terms in the sum for which the near-extremal black hole approximations made above are those with  $Q \gg 1$  and with $\frac{4\pi}{ G_N}\left(\b^2 + \frac{3\b^4}{L^2}\right) \gg Q^2$ (this is equivalent to $\beta \gg r_0(Q)$). Consequently, in order for the sum \eqref{eq:partition-function-sum-over-q} to be valid we need it to be dominated by charges within this (very large) range. This problem is only well-defined when the sum converges, which only happens at finite $L$ (in flat space the integrand grows too fast with the charge). Therefore, when fixing the chemical potential we will only consider finite $L$. 

In order to make contact with previous work in the literature and simplify the equivalent boundary theory, it is interesting to study the dominating charge within this sum and the charge fluctuations around it. 

In the large charge limit the Schwarzian contribution is order one and balancing only the chemical potential and mass term gives
\beq
\partial_Q\Big(\mu\frac{Q}{e}-M_0\Big)\Big|_{Q_0}=0~~~\Rightarrow~~~Q^2_0 = \frac{(4\pi)^2 L^2 \mu^2}{3 e^4} (4\pi G_N \mu^2 - e^2). 
\eeq
The near-extremal approximation is valid as long as $\mu \ll \frac{e}{2L} \sqrt{\frac{L^2+3\b^2}{G_N}}$. 
This formula is consistent with \eqref{rnbmu} but now the extremal charge $Q_0$ should be thought of as a function of $\mu$. This extremal value of the charge is not the true saddle point of the full partition function in \eqref{eq:partition-function-sum-over-q}. It is useful anyways to expand around it $Q=Q_0 +e q$, such that $q\in \mZ$. Then keeping terms up to quadratic order in $q$ we obtain 
\be \label{eq:Zrnsimp}
Z_{\rm RN}[\b, \mu] =e^{\beta \mu \frac{Q_0}{e}} e^{\pi \Phi_0(Q_0)} e^{-\b M_0(Q_0)}  \sum_{q \in  \mZ} e^{2 \pi \mathcal{E} q - \beta \frac{q^2}{2K}}  Z_{\rm Sch}(\Phi_{b,Q_0+eq}, \beta), 
\ee 
where following \cite{Sachdev:2019bjn}, we defined the coefficients 
\beq\label{eqKe}
K \equiv \frac{4\pi(L^2+6 r_0^2)}{3 e^2 r_0}=\frac{4\pi L^2 r_0}{3 e^2 L_2^2}  ,~~~\mathcal{E} \equiv \frac{e L r_0 \sqrt{L^2 +3 r_0^2}}{\sqrt{4 \pi G_N} (L^2+6 r_0^2)}=\frac{L_2^2}{4 \pi }  \frac{Q_0}{r_0^2} \,.
\eeq
It is easy to understand in general the origin of these terms. The chemical potential and mass terms do not produce linear pieces since $Q_0$ is chosen for them to cancel. Then the linear piece in $Q$ comes purely from expanding $ S_0 = \pi \Phi_0(Q_0+e q)$ to linear order. This gives 
\beq
2\pi \mathcal{E} = e\Big(\frac{\partial S_0}{\partial Q}\Big)_{T=0}, 
\eeq
which we can verify also directly from \eqref{eqKe} and matches with Sen's relation between the charge dependence of the extremal entropy and the electric field near the horizon \cite{Sen:2005wa}. A similar argument gives the prefactor of the quadratic piece (coming to leading order from the $\beta\mu \frac{Q}{e}-\beta M$ term) as 
\beq
K=\frac{1}{e}\Big(\frac{\partial Q}{\partial \mu}\Big)_{T=0},
\eeq 
which also is consistent with \eqref{eqKe} and with the results of \cite{Sachdev:2019bjn}. 

The first three terms of \eqref{eq:Zrnsimp} give the extremal contribution to the partition function while the sum includes energy fluctuations (through the Schwarzian) and charge fluctuations. These are not decoupled since the Schwarzian coupling depends on the charge. Nevertheless it is easy to see that corrections from the charge dependence of the dilaton are suppressed in the large $Q_0$ limit and can be neglected (this can be checked directly from \eqref{eq:schwarzianpf}). Then we have
\be 
Z_{\rm RN}[\b, \mu] =e^{\beta \mu \frac{Q_0}{e}} e^{\pi \Phi_0(Q_0)} e^{-\b M_0(Q_0)}  Z_{\rm Sch}(\Phi_{b,Q_0}, \beta)  \sum_{q \in  \mZ} e^{2\pi \mathcal{E} q - \beta \frac{q^2}{2K}} 
\ee 
The partition function in this limit can be reproduced by a one dimensional theory that is a simplified approximation of the one presented in the previous section for small charge fluctuations around the extremal value
\beq
I_{\rm Sch \times U(1)} = \Phi_{b,Q_0} \int_0^\beta du~ {\rm Sch}\Big(\tan \frac{\pi \tau}{\beta},u \Big) + \frac{K}{2} \int_0^\beta du \Big( \theta'(u)+ i \frac{2\pi \mathcal{E}}{\beta} \tau'(u) \Big)^2,
\eeq
written in terms of the field $\tau(u)$. This matches the result of \cite{Sachdev:2019bjn} obtained from a different perspective. As explained in the introduction the main point of this paper is to present a derivation that clarifies the fact that this analysis is true at energies lower than the gap scale. Therefore we conclude that besides matching the semiclassical thermodynamics, the quantum corrections of this theory are also reliable. The exact partition function of the Schwarzian mode was given in \eqref{eq:schwarzianpf} and besides the semiclassical term it only contributes an extra one-loop exact $\frac{3}{2} \log T$ to the partition function. On the other hand the contribution from the $U(1)$ mode is 
\beq
Z_{\rm U(1)}(K,\mathcal{E},\beta) = \sum_{q \in  \mZ} e^{2\pi \mathcal{E} q - \beta \frac{q^2}{2K}}=\theta_3\Big(i\frac{\beta}{2\pi K}, i\mathcal{E} \Big)
\eeq
so the total partition function is given by 
\be 
Z_{\rm RN}[\b, \mu] =e^{\beta \mu \frac{Q_0}{e}+S_0(Q_0)-\b M_0(Q_0)} \Big( \frac{\Phi_{b,Q_0}}{\beta}\Big)^{3/2} e^{\frac{2\pi^2 }{\beta}\Phi_{b,Q_0}} ~\theta_3\Big(i\frac{\beta}{2\pi K}, i\mathcal{E} \Big),
\ee 
where $\theta_3$ is the Jacobi theta function. In this formula $Q_0$ is seen as a function of the chemical potential.

In general we do not need the full result for the U(1) mode. The partition function is dominated by a charge $q= 2\pi K \mathcal{E}/\beta$ giving a saddle point contribution $ \log Z^{\rm s.p.}_{U(1)} = 2 \pi^2 \mathcal{E}^2 K/\beta$. We can define a $U(1)$ scale by\footnote{If we consider a black hole in flat space $L\to\infty$ and $ M_{U(1)}\to 0$, leading to large charge fluctuations. This is related to the fact that the sum over charges is divergent in flat space.}
\beq
M_{U(1)} \equiv 2 K^{-1}=  M_{{\text{\tiny{SL(2)}}}} \frac{3}{2\pi} \frac{ e^2 L_2^4}{L^2 G_N}.
\eeq
For $T\ll M_{U(1)}$ charge fluctuations are frozen since their spectrum does have a gap of order $M_{U(1)}$ and thermal fluctuations are not enough to overcome it. For $T\gg M_{U(1)}$ the $U(1)$ mode becomes semiclassical and its one-loop correction can contribute an extra factor of ${\small \frac{1}{2}}\log T$ to the partition function (see \cite{Mertens:2019tcm} for more details of these limits) from its approximate continuous spectrum. 

For large black holes in AdS, $M_{U(1)}\sim M_{{\text{\tiny{SL(2)}}}} \frac{e^2 L^2}{G_N}$ and therefore is a tunable parameter depending on $e$. If $e$ is small but order one, then $M_{U(1)} \gg M_{{\text{\tiny{SL(2)}}}}$ and for $T\sim M_{{\text{\tiny{SL(2)}}}}$ there is no $\frac{1}{2}\log T$ contribution and charge fluctuations are frozen.  If the theory is supersymmetric then $e^2 \sim G_N$ and $M_{U(1)} \sim M_{{\text{\tiny{SL(2)}}}}$.

\subsection{Density of states at $j=0$}
\label{sec:DOS}
In the previous section we computed the partition function and free energy of the black hole. We can also look at the density of states directly as a function of energy and charge, for states of vanishing angular momentum. For this we can start from \eqref{eq:partition-function-sum-over-q} and solve the Schwarzian theory first. This gives 
\be 
Z_{\rm RN}[\b, \mu] = \sum_{Q \in e \cdot \mZ} e^{\beta \mu \frac{Q}{e}} e^{\pi \Phi_0(Q)} e^{-\b M_0(Q)} \int_0^\infty ds^2~ \sinh(2 \pi s )e^{-\beta \frac{s^2}{2 \Phi_{b,Q}}}
\ee 
This can be used to automatically produced the Legendre transform of the partition function giving the density of states. Now we can define the energy as $E = M_0(Q)+ \frac{s^2}{2 \Phi_{b,Q}}$ to rewrite this expression in a more suggestive way as 
\be 
Z_{\rm RN}[\b, \mu] =   \sum_{Q \in e \cdot \mZ} \int_{M_0(Q)}^\infty dE~ e^{S_0(Q)}\sinh\Big[2 \pi \sqrt{2\Phi_{b,Q}(E-M_0(Q))} \Big] ~e^{\beta \mu \frac{Q}{e}-\beta E}.
\ee 
From this expression we can read off the density of states for each fixed charge $Q$ sector as 
\beq\label{eq:dossinh}
\rho(E,Q)=e^{S_0(Q)}\sinh\Big[2 \pi \sqrt{2\Phi_{b,Q}(E-M_0(Q))} \Big] \Theta(E-M_0(Q)), 
\eeq
where $M_0(Q)$ and $S_0(Q)$ are the mass and entropy associated to an extremal black hole of charge $Q$ while $\Phi_{b,Q} = M_{{\text{\tiny{SL(2)}}}}^{-1}$. At large energies we can match with semiclassical Bekenstein-Hawking expanded around extremality, while for $E\lesssim M_{{\text{\tiny{SL(2)}}}}$ the density of states goes smoothly to zero as $E-M_0(Q) \to 0$. Therefore there is no gap of order $M_{{\text{\tiny{SL(2)}}}}$ in the spectrum. Finally, as we commented above, the path integral over the matter fields can only produce temperature-independent shifts of $S_0$  and $M_0$ in the partition function. This means that the energy dependence of the expression \eqref{eq:dossinh} is reliable in this limit.

This result is not inconsistent with the analysis of Maldacena and Strominger \cite{Maldacena:1997ih}. In that paper, the authors claim the first excited black hole state corresponds to a state with $j=1/2$, with an energy above extremality that coincides with the gap scale. Here, we have shown that a more careful analysis of the Euclidean path integral shows the presence of excited black holes states of energy smaller than $M_{{\text{\tiny{SL(2)}}}}$, and they are all within the $j=0$ sector.

\subsection{The grand canonical ensemble with fixed boundary metric}
\label{sec:large-BHs-AdS4-grand-canonical-ensemble-2}
Finally we will comment on the situation when we fix the full metric at infinity, instead of the angular momentum. For concreteness we will consider the case of a large black hole in AdS$_4$ with $r_0\gg L$. In this case, the dimensional reduction produces a partition function given by \eqref{eq:partition-function-sum-over-q-and-j} setting the SO(3) chemical potential to zero $ \mu_{\text{\tiny{SO(3)}}}\to 0$\footnote{the result for the general case can be found in appendix \ref{app:SO(3)-gauge-field-solutions}.}. This gives 
\be 
\label{eq:partition-function-sum-over-q-and-j2}
Z_{\rm RN}[\b, \mu] = \sum_{Q \in e \cdot \mZ, j \in \mZ} (2j+1)^2 e^{- \beta \mu \frac{Q}{e}} e^{\pi \Phi_0(Q, j)} e^{-\b M_0(Q, j)}  Z_{\rm Sch}(\Phi_{b,Q},\beta)\,.
\ee 
After repeating the analysis of section \ref{sec:large-BHs-AdS4-canonical-ensemble} we can obtain the following expression 
\be 
\label{eq:partition-function-sum-over-q-and-j3}
\hspace{-0.3cm}Z_{\rm RN}[\b, \mu] =e^{\beta \mu \frac{Q_0}{e}+\pi \Phi_0(Q_0)-\b M_0(Q_0)} Z_{\rm Sch}(\Phi_{b,Q},\beta)Z_{\rm U(1)}(K,\mathcal{E},\beta) \sum_{ j \in \mZ} (2j+1)^2 e^{-\beta \frac{G_N j(j+1)}{2r_0^3}} \,.
\ee
Since the correction in the energy from spin $\delta M = G_N j(j+1)/2r_0^3$ is very small for large macroscopic black holes (being suppressed by $G_N$ and also by $r_0$) we can approximate the contribution of the SO(3) gauge field by 
\be 
Z_{\rm RN}[\b, \mu] =e^{\beta \mu \frac{Q_0}{e}+\pi \Phi_0(Q_0)-\b M_0(Q_0)} Z_{\rm Sch}(\Phi_{b,Q},\beta)Z_{\rm U(1)}(K,\mathcal{E},\beta) \Big( \frac{G_N}{2r_0^3\beta} \Big)^{3/2}\,.
\ee
Therefore at low temperatures, $T\ll T_{\rm U(1)}$, the non trivial temperature dependence of the partition functions is given by 
\beq
Z_{\rm RN}[\b, \mu] \sim e^{\beta \mu \frac{Q_0}{e}+S_0-\b M_0} \Big( \frac{\Phi_{b,Q_0}}{\beta}\frac{G_N}{2r_0^3\beta}\Big)^{3/2} e^{\frac{2\pi^2}{\beta}\Phi_{b,Q_0}}\,.
\eeq

As a final comment, in a similar manner to the previous section, we can write a simplified, approximate, one dimensional theory capturing the physics of these states. We need to add an extra term 
\bea
I_{\rm Sch \times U(1)\times SO(3)} &=& \Phi_{b,Q_0} \int_0^\beta du~ {\rm Sch}\Big(\tan \frac{\pi \tau}{\beta},u \Big) + \frac{K}{2} \int_0^\beta du \Big( \theta'+ i \frac{2\pi \mathcal{E}}{\beta} \tau' \Big)^2\nn\\
&&+ 
\frac{K_{\text{\tiny{SO(3)}}}}{2} \int {\rm Tr}\Big[h^{-1} h'+i \frac{\mu_{\text{\tiny{SO(3)}}}}{\beta} \tau' \Big]^2,
\ea
where $K_{\text{\tiny{SO(3)}}}=r_0^3/G_N$. This is a simplification of the more general action written down previously in equation \eqref{eq:particle-moving-on-group-manifold-general} since it only captures fluctuations around the angular momentum saddle-point in the sum \eqref{eq:partition-function-sum-over-q-and-j3}. From the discussion here its clear that the prefactor of the $SO(3)$ action is given by 
\beq
K_{\text{\tiny{SO(3)}}} =\frac{1}{2} \Big(\frac{\partial J^2}{\partial E}\Big)_{T=0}.
\eeq
Finally the gap scale for the $SO(3)$ mode is given by 
\beq
M_{\text{\tiny{SO(3)}}} = 2\frac{G_N}{r_0^3} = M_{{\text{\tiny{SL(2)}}}} \frac{L^2}{r_0^2}\ll M_{{\text{\tiny{SL(2)}}}}\,.
\eeq
Therefore when we fix the boundary metric the sphere modes produce an extra factor of $\frac{3}{2} \log T$ as long as $T\gg M_{\text{\tiny{SO(3)}}}$. For $T\ll M_{\text{\tiny{SO(3)}}}$ the thermal energy is not large enough to overcome the gap of this sector, the angular momentum is frozen, and it does not contribute to $\log{T}$ factors. If we are interested in scales of order, $M_{{\text{\tiny{SL(2)}}}}$ then we are always above the gap for the $SO(3)$ mode.

\section{Contributions from massive Kaluza-Klein modes}
\label{sec:massive-KK-modes}

In the previous section we neglected the contribution from massive Kaluza-Klein modes to the the partition function at low temperatures $T \sim M_{{\text{\tiny{SL(2)}}}}$. We will argue that this is correct in this section. First, we will summarize the spectrum of masses for the remaining Kaluza-Klein modes in the Reissner-Nordstr\"om solution, following the analysis of \cite{Michelson:1999kn}. As an example, we perform the dimensional reduction of the $4d$ scalar field in the theory, to obtain the contribution of the Kaluza-Klein modes to the action of the  $2d$ theory. Then, we will argue that the partition function of massive fields does not contribute to the leading temperature dependence close to extremality.

\subsection{A summary of the Kaluza-Klein spectrum of masses}

The full analysis involving the metric KK modes and the gauge field KK modes is very complicated. Instead, since we will be most interested in the spectrum of masses, a linearized analysis is enough. This was done in detail by Michelson and Spradlin \cite{Michelson:1999kn} (see also \cite{Larsen:2014bqa}). As we will explicitly show for the case of a $4d$ scalar field, the dimensional reduction can be performed by decomposing the fields into scalar or vector spherical harmonics (labeled by the spin $\ell$) on the internal $S^2$ space.
 
At the $\ell=0$ level \cite{Michelson:1999kn} found two relevant modes. One is the dilaton and two-dimensional metric, which combine into JT gravity and also the s-wave of the gauge field, which gives a massless $2d$ U(1) field, as pointed out in section \ref{sec:dimensional-reduction}. At $\ell=1$ level, we have a massive $2d$ scalar and vector coming from the gauge field and a massless field from the metric which coincides with the $2d$ gauge field $B$ related to the $SO(3)$ symmetry of $S^2$ (which we also already identified in \ref{sec:dimensional-reduction}). Finally, for $\ell\geq2$,  \cite{Michelson:1999kn} found massive graviton KK modes (although they point out they are not independent degrees of freedom on-shell) and massive vector degrees of freedom from KK modes of the dilaton and $U(1)$ gauge field. Therefore, besides the massless modes that we have already considered in section \ref{sec:near-extremal-in-AdS4} and \ref{sec:part-function-large-BHs-AdS4}, we solely have massive fields whose minimum mass is given by $m^2 = 1/\chi^2$.

\subsection{An example: the dimensional reduction of a $4d$ scalar}

To clarify the summary, we will give the simplest example of a massive mode appearing in the KK reduction of a scalar field in four dimensions. The action for a scalar field $X$ of mass $m$ is 
\beq
I_{X}=\int d^4x \sqrt{g_4} (g_4^{AB} \partial_A X \partial_B X + m^2 X^2).
\eeq 
In order to carry out the KK reduction we wrote an ansatz for the metric \eqref{eq:beyond-S-wave-metric-ansatz}. To compute the action of the KK modes it is useful to write explicitly the inverse metric in this notation, which is given by 
\beq
g_4^{\mu\nu} = \chi^{1/2} g_2^{\mu\nu},~~~~g_4^{m \mu} = -\chi^{1/2} B^{a\mu} \xi^m_a ,~~~~g_4^{mn} = \frac{1}{\chi} h^{mn} + \chi^{1/2} B^a_\mu B^{b\mu} \xi_a^m \xi_b^n,
\eeq
where the $\mu$ index of $B$ is raised with the $2d$ metric. Also, the determinant of the metric is $g_4 = \chi g_2 h$. We will expand the scalar field into spherical harmonics as 
\beq
X(x,y) = \sum_{\ell} \mathbf{X}_{\ell}(x) \cdot \mathbf{Y}^\ell(y)
\eeq
where we use the (uncommon) notation of denoting the scalar spherical harmonics of spin $\ell$ as a vector $\mathbf{Y}^\ell(y) = [Y^\ell_{-\ell} (y),Y^\ell_{-\ell+1} (y),\ldots, Y^\ell_{\ell} (y)]$. Correspondingly, we denoted the modes of the scalar field also a vector in a similar way $\mathbf{X}_{\ell}(x) =[ X_{\ell}^{-\ell},X_{\ell}^{-\ell+1},\ldots, X_{\ell}^\ell ]$. Then the inner product above denotes $ \mathbf{X}_{\ell}(x) \cdot \mathbf{Y}^\ell(y) \equiv \sum_{m} X_\ell^m Y_\ell^m$.

We will begin by reducing the kinetic term. For this we need the inverse metric and its clear it will produce terms linear and quadratic in the gauge field $B$. The following formulas for integrating spherical harmonics will be useful 
\beq
\int_{S^2} dy \sqrt{h} \mathbf{Y}_\ell^\dagger (\xi_a\cdot \partial) \mathbf{Y}_{\ell'} = i T^a \delta_{\ell' \ell},~~~\int_{S^2} dy \sqrt{h} (\xi_a\cdot \partial)  \mathbf{Y}_\ell^\dagger(\xi_b\cdot \partial) \mathbf{Y}_{\ell'} = - T^aT^b \delta_{\ell' \ell}, 
\eeq
where $\xi_a$ denote the Killing vectors of the sphere and since this is a matrix equation the $T^a$ are matrices giving the spin $\ell$ representation of the rotation group. Then we can obtain the reduction of the kinetic term as 
\beq
\int d^4x \sqrt{g_4} (\partial X)^2 = \sum_\ell  \int d^2x \sqrt{g} \chi^{1/2} \left[ (D_\mu \mathbf{X}_\ell)^\dagger (D^\mu \mathbf{X}_\ell) - \frac{\ell(\ell+1)}{\chi} \mathbf{X}_{\ell}^\dagger \mathbf{X}_{\ell}\right],
\eeq
where we also used the fact that $\Box_{S^2} \mathbf{Y} = - \ell(\ell+1) \mathbf{Y}$, where $\Box_{S^2}$ is the laplacian on the two-sphere. We also defined the covariant derivative 
\beq
D_\mu \mathbf{X}=\partial_\mu \mathbf{X}- i B^a_\mu (T^a)_{\ell} \mathbf{X},
\eeq
where $(T^a)_{\ell}$ are the spin $\ell$ representation matrices acting on the vector $\mathbf{X}$. Adding the mass term, we can obtain the full $2d$ action for the KK reduction of the scalar field as 
\beq
I_X = \sum_\ell \int d^2x\sqrt{g} \chi^{1/2} (| D \mathbf{X}_\ell|^2 - m_\ell^2 |\mathbf{X}_\ell|^2),~~~m_\ell^2 = m^2 + \frac{\ell(\ell+1)}{\chi}.
\eeq
To summarize, a single scalar field KK reduces to a tower of massive fields $\mathbf{X}_\ell$ of dimension $(2\ell+1)$ with $\ell=0,1,2,\ldots$ with increasing mass. 

This is a complicated action: besides being coupled to the two-dimensional metric, it is also coupled to the dilaton. The dilaton coupling is not particularly useful in the FAR region since the dilaton varies with the radius. Of course, in this region, the picture of a single scalar in the $4d$ black hole background is more appropriate. In the NHR this becomes very useful since $\chi \approx \chi_0 = r_0^2$. Then we end up, after rescaling $\mathbf{X}_\ell \to r_0^{-1/2} \mathbf{X}_\ell$ in the NHR with a tower of KK modes with action 
\beq
I_X = \sum_\ell \int d^2x\sqrt{g}(| D \mathbf{X}_\ell|^2 - m_\ell^2 |\mathbf{X}_\ell|^2),~~~m_\ell^2 = m^2 + \frac{\ell(\ell+1)}{r_0^2},
\eeq
fixing the KK mode scale $\Lambda_{\rm KK}\sim 1/r_0$. Naively it seems the correction to the mass is small, but we will take such low temperatures that $\beta \Lambda_{\rm KK} \gg 1$. Then we end up with a tower of canonically normalized free fields. 
 
We can see what happens when turning on scalar field interactions in the initial $4d$ theory. To simplify lets consider self interactions of the scalar field $I_n = \lambda_n \int d^4 x \sqrt{g_4} X^n$.  After KK reducing, this produces a term of order $\lambda_n r_0$. After rescaling the scalar field by $r_0^{-1/2}$ to make the $2d$ action canonically normalized the effective two dimensional coupling becomes $\lambda_n^{\rm 2d} = \lambda_n r_0^{1-n/2}$. Therefore even if selfinteractions are large in four dimensions, the reduction to two dimensions gives $\lambda_n^{\rm 2d} \to 0$ (for large $r_0$) and therefore, in the NHR, its enough to consider free fields. Moreover, since we will only consider states for which fluctuations in the gauge field are small $B \sim j/r_0^3$ we will also neglect its coupling to $2d$ matter.

\subsection{The massive Kaluza-Klein modes in the partition function}\label{sec:KKmodespart}

As we have summarized in the previous subsection, besides the $2d$ massless gravitational and gauge degrees of freedom, all other modes generated by the dimensional reduction have masses given by the value of the dilaton field at the horizon $1/\chi^2$. Furthermore,  as we observed in section \ref{sec:near-extremal-in-AdS4}, the dominating background for the $SO(3)$ gauge fields is that in which they are turned off, $B^a = 0$. Therefore, we will assume that the massive modes are decoupled from the $SO(3)$ gauge field. With this set-up in mind, we can now proceed to compute the contribution to the partition function of the massive KK modes. To show that such fields do not yield any correction to the $\log(T)$ term, we will solely focus on scalar fields and compute their contribution in the NHR. As discussed in preceding subsections, in such a region, their mass is constant and given by $m^2 = 1/r_0^2$. We will also ignore the fluctuations of the Schwarzian boundary mode because the contribution of these fluctuations to the massive modes is suppressed by the scale $\epsilon/r_0$ from \eqref{eq:AdS2-scales-and-bdy-cond}.  

Therefore, we will compute the contribution of the massive modes in a circular patch of the Poincar\'e disk, where the proper length of the boundary is $\ell = \b L_2/\e$ and its extrinsic curvature is constant. We will choose Dirichlet boundary conditions for the scalar field $\bX|_{\partial M_{\rm NHR}} = 0$ at the boundary $\partial M_{\rm NHR}$; this is consistent with the classical solution $\bX$ for the field in the FAR when fixing $\bX|_{M_2} = 0$. The contribution of a KK mode in the NHR is then abstractly given by $Z_{KK} = \det(g_{\mu \nu}^{\rm NHR} \partial^\mu \partial^\nu + r_0^{-2})^{-1/2}$. 

To compute the $\beta$-dependence of this determinant we will us the Gelfand-Yaglom method \cite{gel1960integration}, studying the assymptotics of solutions to the Klein-Gordon equation $(\Box_{\rm NHR}+m^2) \psi = 0$.\footnote{This strategy was previously used to study the mass-dependence of the determinant \cite{Maldacena:2019cbz}.}  Parametrizing the $AdS_2$ coordinates by $ds_{\rm NHR}^2   =L_2^2\left(dr^2 + \sinh^2(r) d\phi^2\right)$, we find that the boundary is located at $r_{\partial {\rm NHR}} = \log\left(\frac{\b}{\pi \e}\right)   + O(\b^2/\e^2) \rightarrow \oo $.  Expanding $\psi(r, \phi) = \psi_k(r) e^{ik\phi}$ with $k \in \mZ$, the Klein-Gordon equation becomes
\be
\label{eq:Klein-Gordon-in-k} 
\frac{1}{\sinh r} \partial_r (\sinh r \partial_r \psi_k) - \frac{k^2}{\sinh^2 r} \psi_k  + (mL_2)^2 \psi_k = 0\,,
\ee
whose regular solution at the horizon ($r=0$) is given by \footnote{The other solution diverges at the horizon. }
\be
\label{eq:contribution-of-the-partition-function}
\psi_k(r) =  \frac{(\tanh r)^{|k|}}{(\cosh \, r)^{\De_+}}\,_2F_1\left(\frac{1}4 + \frac{|k|}2 + \frac{\nu}2,\, \frac{3}4 + \frac{|k|}2 +\frac{\nu}2, 1+|k|, \tanh^2 r \right)\,,
\ee
where we define
\be
 \De_\pm \equiv \frac{1}2 \pm \sqrt{\frac{1}4 + (mL_2)^2} \,, \qquad \nu = \sqrt{\frac{1}4 + (mL_2)^2}\,.
\ee
The Gelfand-Yaglom method requires that we normalize $\psi_k$ such that its derivative at $r=0$ is independent of $m$; this is indeed the case, when expanding \eqref{eq:contribution-of-the-partition-function} to first order in $r$ around the horizon. 
Asymptotically, for $r = r_{\partial {\rm NHR}}\rightarrow \infty$, the solution is given by 
\be 
\hspace{-0.5cm}\psi_k  = \frac{\Gamma(1+|k|)2^{|k|}}{\sqrt{\pi}} \left[\frac{1}{(2\cosh r_{\partial {\rm NHR}})^{\De_-}} \frac{\Gamma(\De_+-1/2)}{\Gamma(\De_+ +|k|)} + \frac{1}{(2\cosh r_{\partial {\rm NHR}})^{\De_+}} \frac{\Gamma(\De_--1/2)}{\Gamma(\De_- +|k|)}\right]\,.
\ee
The Gelfand-Yaglom theorem states that the determinant with Dirichlet boundary conditions for the scalar field is given by $\det(\Box_{\rm NHR}+m^2) = \cN(\b, \e) \prod_k \psi_k(r_{\partial {\rm NHR}})$, where $\cN(\b, \e)$ is a mass-independent proportionality constant. 

The contribution to the free energy coming from the determinant is then given by, 
\be
\label{eq:KK-reduction}
\log Z_{KK} = -\frac{1}{2}\log\cN(\b, \e)-\frac{1}{2}\sum_{k \in \mZ} \log \left[\frac{\Gamma(1+|k|)2^{|k|}}{\sqrt{\pi} (2\cosh r_{\partial {\rm NHR}})^{\De_-}} \frac{\Gamma(\De_+ - 1/2)}{\Gamma(\De_+ + |k|)}\right]\,.
\ee
To determine $\cN(\b, \e)$ we use the result for the partition function of a massless scalar on a circular patch of the Poincar\'e disk \cite{Yang:2018gdb}. Since the massless scalar can be treated as a $2d$ CFT,  the result can be determined by computing the Weyl anomaly when mapping a unit-disk in flat-space to the circular $AdS_2$ patch of interest.  The first few orders in the large $\b$ expansion of the free energy obtained from the Weyl-anomaly are given by,
\be 
\label{eq:massless-part-function}
\log Z_{m^2 = 0}  = \frac{c}{24} \frac{\b}{\pi \e}+ \frac{c}6 \Big[ \log{(2L_2)}-\frac{1}{2}\Big] + O\Big(\frac{\e}{\b}\Big)\,,
\ee
where $c = 1$ is the central charge of one free boson \footnote{To get this result, we write the metric of the hyperbolic disk at finite cut-off $g$ as $g=e^{2\rho}\hat{g}$ where $\hat{g}$ is the flat unit disk metric. Then we evaluate the Liouville action for the particular choice of $\rho$ associated to the hyperbolic disk and expand for small $\e$.}. The term at order $O(\b/\e)$ can in principle be canceled by adding a cosmological constant counter-term to the boundary of the NHR, $I_{\text{counter-term},\,CFT} = \int_0^\b du \,\,c \sqrt{h_{uu}}/(24\pi)$. However, since we are solely interested in reproducing the $\log \b$ dependence of the free energy we will not delve into how this term is reproduced by studying the coupling of these scalars to the FAR. 

At such low temperatures, the Schwarzian mode is strongly coupled, so we might be worried that it can affect the answer. In \cite{Yang:2018gdb} it was observed that the boundary Schwarzian fluctuations lead to correction of $O(\e)$ to the partition function \eqref{eq:massless-part-function}. Since we expect the same to be true when turning on a mass, the contribution of the Schwarzian fluctuations to the partition function of the Kaluza-Klein fields can be safely ignored.

 Therefore, up to terms proportional to $\b/\e$ obtained from the counter-term, this fixes  
\beq
\label{eq:partition-Z-kk}
\log Z_{KK} = \frac{1}6 \Big[ \log{(2L_2)}-\frac{1}{2}\Big]  -\frac{1}2 \sum_{k \in \mZ} \log \left[\frac{1}{(2\cosh r_{\partial {\rm NHR}})^{\De_-}} \frac{\Gamma(1+|k|)}{\Gamma(\De_+ + |k|)}\right]\,.
\eeq
The sum in \eqref{eq:KK-reduction} needs to be regularized in order for it to converge; in principle, this can be done by accounting for the divergent non-universal terms in the massless partition function \eqref{eq:massless-part-function}.  The $\b$-dependent factor in the sum appears through the relation $r_{\partial {\rm NHR}}=   \log\left(\frac{\b}{\pi \e}\right) $; consequently,  the sum is given by $-\sum_{k \in \mZ} \Delta_-\log(\cosh r_{\partial {\rm NHR}})  $ which vanishes in $\zeta$-function regularization. Therefore, the contribution of the KK-modes to the partition function is given by
\be 
\log Z_{KK} &=\frac{1}6 \Big[ \log{(2L_2)}-\frac{1}{2}\Big]  -\frac{1}2 \sum_{k \in \mZ} \log \frac{\Gamma(1+|k|)}{\Gamma(\De_+ + |k|)}\,, 
\ee
which, to leading order, is $\b$-independent. In conclusion, to leading order in $O(1/\Phi_0)$, the KK modes only affect the entropy of the black hole and not the shape of the density of states. Consequently, they also to do not change our prior conclusion about the absence of near-extremal black hole gap.

Finally, we will quickly go over a more direct (yet less rigurous) method to compute the functional determinant following \cite{Banerjee:2010qc} \footnote{We would like to thank A. Castro for discussions about the relation between the calculation in this paper and the previously studied $\log{A}$ terms \cite{Banerjee:2010qc, Banerjee:2011jp, Sen:2011ba, Sen:2012cj}.}. The starting point is again $ds_{\rm NHR}^2   =L_2^2\left(dr^2 + \sinh^2(r) d\phi^2\right)$ with a cutoff at $r_{\partial {\rm NHR}}$ (for simplicity we turn off the Schwarzian mode). We will first take the large cut-off limit for the matter fields and impose $\psi \sim (\cosh{r_{\partial{\rm NHR}}})^{-\Delta_+}$ giving eigenvalues that depend only on $L_2$. Then the contribution from the matter field to the partition function is \cite{Banerjee:2010qc} 
\beq\label{eq:heatkernel}
\log Z_{\rm matter} = (\cosh r_{\partial {\rm NHR}}-1) \int_{\epsilon_{\rm UV}}^\infty ds \frac{1}{s} \int_0^\infty d\lambda (\lambda \tanh \pi \lambda) e^{-s\big[\frac{\lambda^2 + \frac{1}{4}}{L_2^2}+m^2\big]} . 
\eeq
 The whole temperature dependence comes then from the prefactor through $\sinh(r_{\partial{\rm NHR}}) = \frac{\beta}{2\pi \epsilon}$ and this is true regardless of the mass. Expanding at large $r_{\partial{\rm NHR}}$ gives 
\beq
\cosh( r_{\partial {\rm NHR}})-1 = \frac{\beta}{2\pi \epsilon} -1 + \mathcal{O}(\epsilon).
\eeq
From this expression we can easily see the matter contribution is only a shift of the extremal mass, or a temperature independent ($L_2$ dependent) finite correction to the partition function which potentially can only correct $S_0$. Following \cite{Banerjee:2010qc} one could even resum the whole tower of KK modes and reach the same conclusion. One might wonder whether imposing boundary conditions for $\psi$ at a finite cut-off might affect the temperature dependence. However, we have already checked through the Gelfand-Yaglom theorem that this does not happen. 

\section{Discussion}
\label{sec:conclusion}

In this paper, we have computed the partition function of $4d$ near-extremal charged and of slowly-spinning black holes, in the canonical and grand canonical ensembles. By showing that we can reliably neglect all massive Kaluza-Klein modes and by solving the path integral for the remaining massless mode in the near-horizon region, we have shown that our result can be trusted down to low-temperatures, smaller than the scale $\sim M_{{\text{\tiny{SL(2)}}}}$. At this energy scale, we find a continuum of states, disproving the conjecture that near-extremal black holes exhibit a mass gap of order $M_{{\text{\tiny{SL(2)}}}}$ above the extremal state. The existence of a continuum of states suggests that the degeneracy of the extremal state is not given by the naive extremal entropy, fixed by the horizon area. Instead, the horizon area fixes the scaling of the density of states and the level spacing of the states. However, as we will discuss in the following subsection, to make a quantitative statement about the scale of this extremal degeneracy, we need to discuss possible non-perturbative contributions to the $2d$ path integral. 

The process of solving the path integral for the massless modes in the $2d$ dimensionally reduced theory, involved obtaining an equivalent $1d$ theory which can be thought to live on a curve at the boundary of the throat, between the near-horizon region and the far-away region. This equivalent $1d$ theory is given by the Schwarzian coupled to a particle moving on a $U(1) \times SO(3)$ group  manifold. Generally, the potential of the particle moving on the $U(1) \times SO(3)$ is quite complicated. However, when looking at the theory that approximates the charge and angular momentum fluctuations in the grand canonical ensemble for black holes in AdS$_4$, the theory is simply given by: 
\beq
\label{eq:SchU(1)SO(3)-final}
I_{\rm Sch \times U(1)\times SO(3)} = I_{\rm Sch}[\tau]+ I_{\rm U(1)}[\theta,\tau] +I_{\rm SO(3)} [h,\tau]
\eeq
where we defined the Schwarzian, $U(1)$ and $SO(3)$ contributions of the action as 
\bea
I_{\rm Sch}[\tau] &=& \frac{1}{M_{{\text{\tiny{SL(2)}}}}} \int_0^\beta du~ {\rm Sch}\Big(\tan \frac{\pi \tau}{\beta},u \Big), \\
I_{\rm U(1)}[\theta,\tau]& =& \frac{1}{M_{{\text{\tiny{U(1)}}}}} \int_0^\beta du \Big( \theta'+ i \frac{2\pi \mathcal{E}}{\beta} \tau' \Big)^2,\\
I_{\rm SO(3)} [h,\tau] &=& \frac{1}{M_{{\text{\tiny{SO(3)}}}}}\int_0^\b du\hspace{0.1cm} {\rm Tr}\Big(h^{-1} h'+i \frac{\mu_{\text{\tiny{SO(3)}}}}{\beta} \tau' \Big)^2,
\ea
where $\theta(u)$ is a compact scalar and $h(u)$ is an element of $SO(3)$ and the mass scales $M_{{\text{\tiny{SL(2)}}}}$, $M_{{\text{\tiny{U(1)}}}}$ and $M_{{\text{\tiny{SO(3)}}}}$ are fixed by thermodynamic relations. Additionally, $M_{{\text{\tiny{SL(2)}}}}$, $M_{{\text{\tiny{U(1)}}}}$ and $M_{{\text{\tiny{SO(3)}}}}$ can be viewed as the breaking scales for each of their associated symmetries ($SL(2, \mR)$,  $U(1)$ and, respectively, $SO(3)$) for the near-horizon region of an ensemble of near-extremal black holes. 

Beyond the goal of resolving the mass-gap puzzle for near-extremal Reissner-Nordstr\"{o}m black holes, the effective $2d$ dimensionally reduced theory of dilaton gravity (and its equivalent boundary theory) provides a proper framework to resolve several future questions, some of which we discuss below. 

\subsection{Other black holes and different matter contents}
\label{sec:black-holes-with-diff-matter}

While we have successfully analyzed the case of Kerr-Newman black holes with small spin, for which we could neglect the sourcing of massive Kaluza-Klein modes for some of the metric components, it would be instructive to compute the partition function of Kerr-Newman black holes for arbitrary spin. An effective $1d$ boundary theory capturing the dynamics of such black holes was recently described in \cite{Anninos:2017cnw, Moitra:2019bub, Castro:2019crn}; however, the quantum fluctuations relevant for understanding the mass-gap puzzle were not analyzed. In the framework described above, resolving such a puzzle for Kerr-Newman black holes amounts to studying how the massive Kaluza-Klein modes are sourced and whether their fluctuations could significantly affect the partition function. If the analysis in section \ref{sec:massive-KK-modes} follows even in when such fields have a non-trivial classical saddle-point,  then it is likely that near-extremal Kerr-Newman black holes do not exhibit a gap for arbitrary angular momenta. 

Perhaps an even more intriguing case is that of near-extremal (and, at the same time, near-BPS) black holes in $4d$ $\cN = 2$ supergravity. As mentioned in the introduction, in such cases, microscopic string theory constructions \cite{Callan:1996dv, Maldacena:1996ds} suggest that the scale $M_{{\text{\tiny{SL(2)}}}}$ should genuinely be identified as the gap scale in the spectrum of near-extremal black holes masses. While an analysis of the proper effective theory describing such black holes is underway \cite{HITZ}, perhaps some intuition can be gained by looking at a related theory that has less supersymmetry: the $\cN = 2$ super-Schwarzian. In such a theory, the partition function was computed \cite{Stanford:2017thb, Mertens:2017mtv} and its resulting spectrum indeed exhibits a gap whose scale is fixed by the inverse of the super-Schwarzian coupling. Since the inverse of the super-Schwarzian coupling coincides with the conjectured gap \cite{Preskill:1991tb, Page:2000dk, Almheiri:2016fws}, it is tantalizing to believe that the thermodynamic mass-gap observed in \cite{Callan:1996dv, Maldacena:1996ds}  is indeed an artifact of supersymmetry \footnote{The exact density of states of the $\cN = 2$ Schwarzian presents a delta function at extremality with weight $e^{S_0}$ which would be consistent with a highly degenerate extremal black hole. This degeneracy is also consistent with previous microscopic counting and shows that it also relies on supersymmetry to work.}. 

It would also be interesting to study the contribution of charged scalar or fermionic fields to the partition function of the near-extremal Reissner-Nordstr\"{o}m  black holes. In AdS, the presence of such fields has been widely used to study the holographic dual for several phases of matter \cite{Gubser:2008px, Gubser:2008wv, Liu:2009dm, Faulkner:2009wj}. For black holes in flat space, it would be nice to compute the contribution from charged matter with $q/m>1$ and see its effect at the level of the microstates. 

Finally, it would be interesting to consider black holes in AdS$_D$, which have known CFT duals. The result of this paper can be interpreted as a universality of their spectrum when looking at large charges and low temperatures. Those degrees of freedom should be properly described by the effective theory found in this paper. One approach to this problem can be to apply the conformal bootstrap at large charge for higher dimensional CFT (this was done for the case of rotating BTZ in \cite{Ghosh:2019rcj}). Another, perhaps more ambitious, approach is to start directly with the boundary theory and try to derive an equivalent quantum mechanical system in the extremal limit. Such a theory would be similar to SYK (would reduce to the Schwarzian and be maximally chaotic) but would be dual to a local bulk (as opposed to also other higher dimensional versions of SYK \cite{Berkooz:2016cvq, Turiaci:2017zwd, Murugan:2017eto}).  

\subsection{Non perturbative effects}
\label{sec:non-pert-effects}

It was recently made precise how including non-trivial topologies in the Euclidean path integral of $2d$ dilaton gravity can fix certain problems with unitarity \cite{Saad:2019lba} (the price to pay when accounting for such non-trivial topologies is to allow for disorder in the boundary theory). In the case of JT gravity the non-perturbative completion is given by a random matrix and one has to sum over all two-dimensional topologies consistent with the boundary conditions. It would be tempting to trust these corrections in the context of a near-extremal black hole. Then the spectrum would be random, with an averaged level spacing of order $e^{-S_0}$ and a non-degenerate ground state (moreover there is an exponentially suppressed probability of lying below the extremality bound, but this can be avoided by considering supersymmetry).

Of course, this is too optimistic in the case of $4d$ near-extremal black holes. Other non-perturbative effects can appear from the $4d$ perspective, which are not captured by JT gravity. For example, one can consider multi-black hole solutions \cite{Maldacena:1998uz} or topology changes that involve the whole $4D$ space. 

Even within JT gravity, there can be configurations with conical defects in two dimensions, which are smooth when uplifted to the higher dimensional metric. These can be important and hint into solving problems with pure $3d$ gravity \cite{Maxfield:2020ale}. For near-extremal black holes in higher dimensions, one would need to include similar geometries.

\subsection{The replica ensemble and the Page curve}

 A procedure was recently found to reproduce the Page curve from the gravitational Euclidean path integral in JT gravity. In order to reproduce the Page curve \cite{Almheiri:2019qdq, Penington:2019kki} computed the radiation Renyi entropy, including replica wormholes. In those calculations, one couples JT gravity in AdS$_2$ with a bath in flat space, making the evaporation of the black hole possible. This setup can be directly understood as an approximate description of an evaporating near-extremal black hole in four dimensions (we can consider this at temperatures $T\gg M_{{\text{\tiny{SL(2)}}}}$ to simplify the problem so that backreaction around each semiclassical saddle is suppressed). 

To turn the recent calculations into a justified approximation, we have to make the following changes. First, the gravitational part of the theory should be JT gravity coupled to the appropriate gauge fields (both KK and the ones sourcing extremality) and coupled to a matter CFT. This theory should then be glued to the $2d$ s-wave reduction of the four-dimensional extremal black hole metric in asymptotically flat space (we assume in this region gravity is weak). This is justified as long as the dominant evaporation channel happens through s-waves (if higher angular momenta are exponentially suppressed). Since this is usually the case,  the calculation of \cite{Almheiri:2019qdq} can be repeated in the context of $4d$ near-extremal black holes. The main complication is to account for the contribution from all the matter fields in this new geometry, and we hope to address this in more detail in future work.

\subsection*{Acknowledgements}

We thank A.~Castro, M.~Heydeman, G.~Horowitz, H.~Lin, D.~Jafferis, J.~Maldacena, H.~Maxfield, S.~Pufu, H.~Verlinde, Y.~Wang, Z.~Yang and W.~Zhao for valuable discussions. We would like to especially thank J.~Maldacena for crucial discussions at early stages of this work. LVI is supported in part by the US NSF under Grant No. PHY-1820651 and by the Simons Foundation Grant No. 488653. GJT is supported by a Fundamental Physics Fellowship.

\appendix

\section{Gravitational Interpretation of the $SO(3)$ gauge fields}
\label{app:SO(3)-gauge-field-review}

\subsection{The Kerr-Newman solution}
When reducing the Einstein action in four dimensions to two dimensions a $SO(3)$ gauge field emerges from the symmetries of the transverse sphere $S^2$. We denoted the charges associated to this field by $J$. In this appendix we will explicitly check that two dimensional solutions with charge $J$ can be uplifted to KN solution in four dimensions. In the approximation where all $SO(3)$ charged fields can be neglected, the angular momentum $J$ on the black hole is directly related to the value of the $SO(3)$ field strength given by the $SO(3)$ Casimir \cite{Anninos:2017cnw}. 

The KN solution in AdS$_4$ with radius $L$ is given by
\be
\label{metext-KN-BHs-AdS}
(ds^{KN})^2={\rho^2\Delta_{\tilde r} \Delta_\theta \over \Sigma} d\tilde \tau^2 + {\rho^2\over \Delta_{\tilde r}} d\tilde r^2 +  \frac{\rho^2}{\Delta_{\tilde \theta}} d\tilde \theta^2 +  \sin^2\tilde \theta {\Sigma\over \rho^2 \Xi^2}(d \tilde \phi + \cB_{\tilde \tau} \, d\tilde t)^2,
\ee
where the mass, angular momentum and charge are parametrized as 
\beq
M = \frac{m}{G_{N} \Xi^2}\,,~~~~J = \frac{ma}{G_{N} \Xi^2}\,, ~~~~Q = \frac{q}{\Xi}\,,~~~~\Xi \equiv 1 - \frac{a^2}{L^2}\,,
\eeq
and the functions appearing in the metric are 
\be
\rho^2 &= \tilde r^2 + a^2 \cos^2 \tilde \theta\,, ~~~~\Delta_{\tilde r} = (\tilde r^2 + a^2 ) \left(  1 + \frac{\tilde r^2}{L^2} \right) - 2   m \tilde r +q^2\,,~~~~\Delta_{\tilde \theta} = 1 - \frac{a^2}{L^2} \cos^2 \tilde \theta\, , \nn\\
\Sigma &= (\tilde r^2 + a^2)^2 \Delta_{\tilde \theta} - a^2 \Delta_{\tilde r} \sin^2 \tilde \theta\,,~~~~\cB_{\tilde \tau} = i \frac{a \Xi [ (a^2 +\tilde r^2) \Delta_{\tilde \theta} -\Delta_{\tilde r}]}{\Sigma}\,. \label{om4}
\ee
For small $a$ the relation between the angular momentum is given (to first order) by $J = M a$. At small $a$ the metric \eqref{metext-KN-BHs-AdS} can be seen as a deformation of the RN solution from \eqref{eq:RN-black-hole-AdS} in which one turns on a non-trivial profile for the $SO(3)$ gauge field with
\be 
\label{eq:deformation-KN-BH}
\delta g_{\mu \nu} dx^\mu dx^\nu = \frac{2\cB_{\tilde \tau} \Sigma \sin^2 \tilde \theta}{\rho^2 \Xi^2} d\tilde \phi d\tilde \tau = 2i a \sin^2 \tilde  \theta (1- f(\tilde r)) d\phi d\tilde \tau ,
\ee
where $f(\tilde r)$ is the function appearing in equation \eqref{eq:RN-black-hole-AdS}. 
 
As we will show the deformation in \eqref{eq:deformation-KN-BH} does not precisely match with the solution for the $SO(3)$ gauge fields inserted into the dimensional reduction ansatz \eqref{eq:beyond-S-wave-metric-ansatz}. Nevertheless, as we will explain in the next subsection, the perturbed solution  for the KN metric $g^{KN}_{\mu\nu} = g^{RN}_{\mu \nu} + \delta g_{\mu \nu}$ will turn to be equivalent, up to  diffeomorphisms, with the  solution for the $SO(3)$ gauge fields inserted into the dimensional reduction ansatz. 

 Thus, to first order at small $J$ (or equivalently in small $a$), the partition function is well approximated by considering the quantization of the $SO(3)$ gauge field coupled to the standard RN metric given in each sector with fixed $Q$. In the next subsections we further show that this approximation is valid by studying the solutions to the equations of motion for the $SO(3)$ gauge field. Furthermore, we show that the average value of angular momentum contributing to the grand canonical partition function does not strongly backreact on the metric (i.e.~its contribution is much smaller than that of the $U(1)$ charge). 

\subsection{Classical $SO(3)$ gauge field configurations}
\label{app:SO(3)-gauge-field-solutions}
In order to compare the perturbed RN solution to the ansatz for the dimensional reduction \eqref{eq:beyond-S-wave-metric-ansatz} we need to solve the equations of motion for the $2d$ $SO(3)$ gauge fields whose contribution to the action is given by \eqref{eq:action-dimensionally-reduced-final},
\be
\label{eq:SO3-contribution}
I_{EM}^{SO(3)} = -\frac{1}{12 G_N r_0} \int_{M_4} \sqrt{g} \chi^{5/2} \Tr(H_{\mu \nu} H^{\mu \nu})\,. 
\ee

We first start with the case in which we fix the boundary holonomy of the $SO(3)$ gauge fields (which corresponds to fixing the boundary metric on $\partial M_4$) rather than the overall charge of the system.\footnote{We thank Silviu Pufu and Yifan Wang for sharing notes during a past project about instanton solutions in $2d$ $SO(3)$ Yang-Mills theory. } For practical purposes, it proves convenient to choose the boundary component of the gauge field to be constant with $B|_{\partial \cM_2} = i \frac{\mu_{\text{\tiny{SO(3)}}}}{\b} T^3 d\tau $ such that the holonomy is given by $\exp(\oint_{\partial \cM_2} B ) = \exp(i \mu_{\text{\tiny{SO(3)}}} \sigma^3)$ with $\mu_{\text{\tiny{SO(3)}}}  \in [0, 2\pi)$ (according to our conventions $T^a = \frac{1}{2} \sigma^a$ with $\sigma$ the Pauli matrices).

 We can find the solution in the gauge in which $B_r = 0$ and make the ansatz that $B = i \frac{\mu_{\text{\tiny{SO(3)}}} T^3}{\b} \xi(r) d\tau$ for some function $\xi(r)$ satisfying $\xi(r_{\partial M_2}) = 1$.  Then, the field strength is $H = i \frac{ \mu_{\text{\tiny{SO(3)}}} T^3}{\b} \partial_r \xi(r) dr \wedge d\tau$ and the equation of motion $d^*H  =0$ implies that $\left[\xi'(r)/\sqrt{g}\chi^{-5/2}\right]' = 0$. Using the solution $\chi(r)  = r^2$, this implies that 
\begin{eqnarray}
\label{eq:metric-deformation} 
&&\xi(r) = \a_1  + \frac{\a_2}{r^3}\,,\qquad  H_{r \tau}  =   - i \frac{ \mu_{\text{\tiny{SO(3)}}}}{\beta} \frac{3\a_2}{r^4}T^3 \nn\\
&& \delta g_{\mu \nu}^{SO(3)} dx^\mu dx^\nu = 2r^2 i \sin^2(\theta) \frac{ \mu_{\text{\tiny{SO(3)}}}}{\beta}\left( \a_1  + \frac{\a_2}{r^3}\right) d\tau d\phi\,.
\end{eqnarray}
Demanding that the gauge field has unit holonomy around the point with $r = r_0$ imposes that $\mu_{\text{\tiny{SO(3)}}} \left(\a_1 + {\a_2}/{r_0^3}\right)=2\pi n$ with $n \in \mathbb Z$. Furthermore imposing that $\xi(r|_{\partial M_2}) = 1$ implies that $\a_1 = 1$, and, consequently, $\a_2 = r_0^3\big(\frac{2\pi n}{ \mu_{\text{\tiny{SO(3)}}}} - 1\big)$. Consequently, we have that 
\begin{eqnarray} 
\label{eq:SO3-instanton-solutions}
B_\tau &=& \frac{i T^3}{\beta} \left[ \mu_{\text{\tiny{SO(3)}}}+ \frac{r_0^3}{r^3}\left(2\pi n -  \mu_{\text{\tiny{SO(3)}}}\right) \right]\nn\\
\delta g_{\mu \nu}^{SO(3)} dx^\mu dx^\nu &=& 2i r^2\frac{\sin^2( \theta)}{\beta} \left[\mu_{\text{\tiny{SO(3)}}}+ \frac{r_0^3}{r^3}\left(2\pi n -  \mu_{\text{\tiny{SO(3)}}}\right) \right]d\tau d\phi\,.
\end{eqnarray}
Gauge field configurations with different $n$ correspond to different instanton configurations for the $SO(3)$ gauge field and different metric solutions, all obeying the same boundary condition on $\partial M_2$.  

As a consistency check, when adding the metric defomation in \eqref{eq:metric-deformation} to the RN metric as in the ansatz \eqref{eq:beyond-S-wave-metric-ansatz} Einstein's field equations are still satisfied to first order in an expansion in $1/\beta$, meaning that the action \eqref{eq:action-dimensionally-reduced-final} resulting from the dimensional reduction is correct. The total action \eqref{eq:metric-deformation} evaluates to
\begin{eqnarray}
\label{eq:total-action-eval}
 H^3_{r \tau}  &=&   - i \frac{1}{\beta} \frac{3r_0^3}{r^4}\left(2\pi n-\mu_{\text{\tiny{SO(3)}}} \right)\,,\nn\\
  \qquad I_{EM}^{SO(3)} &=& \frac{1}{6 G_N}\int_0^\b  d\tau  \int_{r_0}^\infty dr \frac{9r_0^6}{\beta^2 r^4} (2\pi n-\mu_{\text{\tiny{SO(3)}}})^2 = \frac{2r_0^3 (2\pi n- \mu_{\text{\tiny{SO(3)}}})^2}{G_N \beta}
\end{eqnarray}
in each instanton sector. To make contact with the effective action in each $j$ sector in the sum over $SO(3)$ representations we can evaluate the sum over $j$ for the contribution of each representation to the partition function \eqref{eq:action-Q,J-sector} for the on-shell solution $\chi(r) = r^2$:
\be
\label{eq:rep-to-SO3=instanton} 
Z^{SO(3)}_{RN} &= 
\sum_{j \geq 0} (2j +1) \chi_{j}(\mu_{\text{\tiny{SO(3)}}}) e^{-\frac{ G_N \beta}{2 r_0^3}  j(j+1)}= - \frac{e^{\frac{G_N \b}{8 r_0^3}}}{4 \sin \mu_{\text{\tiny{SO(3)}}}} \vartheta_3' \Big(\zeta/2, e^{-\frac{G_N \b}{8 r_0^3}} \Big)  \nn\\  &= \sum_{n \in \mathbb Z} \frac{2 \sqrt{ \pi} (\mu_{\text{\tiny{SO(3)}}} - 2 \pi n) }{\big(\frac{G_N \b}{2r_0^3}\big)^{3/2}  \sin (\mu_{\text{\tiny{SO(3)}}} - 2 \pi n) } e^{-\frac{2r_0^3}{G_N \b}(\mu_{\text{\tiny{SO(3)}}} - 2 \pi n)^2}\,,
\ee
where  $\vartheta_3'(u, q)$ is the derivative with respect to $u$ of  $\vartheta_3(u, q)$ and where to obtain the final equation we have used the expansion in terms $\frac{G_N \b}{r_0^3}$. Consequently, we find that the sum over instanton saddle in the partition function \eqref{eq:total-action-eval} precisely agrees with the sum over $SO(3)$ representations appearing in the partition function associated to the action \eqref{eq:action-Q,J-sector}.\footnote{The prefator in front of the exponent in \eqref{eq:rep-to-SO3=instanton} can in fact be obtained by computing the one-loop correction to each instanton saddle. The fact that the one loop expansion recovers the complete result is related to the fact that the path integral in $2d$ Yang-Mills theory can be obtained using localization techniques.    } 

To find the relation between the $SO(3)$ representation $j$ and the angular momentum it proves convenient to also analyze the classical solutions in the case in which we fix the field strength at the boundary (or equivalently the Lagrange multiplier zero-form $\phi^{SO(3)}$). In this case, we will fix gauge such that  $H_{r\tau} dr \wedge d\tau|_{\partial M_2} = i\sqrt{g} \sigma_3 \frac{3G_N r_ 0}{\sqrt{2} \chi^{5/2}} j |_{\partial M_2}  =i \frac{3G_N \sigma_3 j}{\sqrt{2}r^4}|_{\partial M_2}  $, for some constant $j$. The resulting gauge field, field strength and $4d$ metric perturbation is given by  
\be
\label{eq:SO3-gauge-field-sol-fixedF}
& B_\tau = i T^3 \left(\frac{\sqrt{2}G_N j}{r_h^3} + \frac{2\pi n}{\b}- \frac{\sqrt{2}G_N j}{r^3}\right)\,, \qquad H_{r\tau}  = i \frac{3G_N \sigma_3 j}{\sqrt{2}r^4}\,, \nn  \\ &\delta g_{\mu \nu}^{SO(3)} dx^\mu dx^\nu = 4 i r^2 \sin^2(\theta) \left[\frac{G_N j}{\sqrt{2}r_0^3} + \frac{2\pi n}{\b}- \frac{G_N j}{\sqrt{2}r^3}\right] d\tau d\phi \,,
\ee
where we have fixed gauge such that $B_r = 0$ and have once again obtained the first $r$-independent term in $B_\tau$ by requiring unit holonomy around the point with $r =r_h$ (i.e.~nowhere is $H$ singular). 

Next we determine the contribution of the $SO(3)$ gauge field to the action. As for the $U(1)$ gauge field \eqref{eq:boundary-term-fixed-F}, in order for to have a well defined variational principle we need to add a boundary term to the action: $I_{EM}^{SO(3), N} = I_{EM}^{SO(3)} + \frac{1}{12G_N r_0}\int du \sqrt{g}\, n_{\mu} \tr H^{\mu \nu} B_{\nu} $. Accounting for this boundary term we find that the 
\be
\label{eq:Newman-bc-action-result}
 I_{EM}^{SO(3), N} = \frac{1}{6} \int_0^\b d\tau \int_{r_0}^\infty dr  \frac{G_N j^2}{r^4} - \int_0^\b d\tau \left(\frac{G_N j}{3r_0^3} + \frac{2\pi n}{\b}\right) j \sim \frac{G_N\beta  j^2}{r_0^3} + 2\pi n j \,.
\ee
 We need to be careful about the $n$-dependent term appearing in the final result in \eqref{eq:Newman-bc-action-result}. If the solutions \eqref{eq:Newman-bc-action-result} are gauge inequivalent then, in order to obtain the partition function, we truly have to sum over all different instanton solutions; since the sum over $n$ is unbounded, the partition function would be ill defined. Consequently, the only possibility is that the gauge field solutions in \eqref{eq:SO3-gauge-field-sol-fixedF} are in fact all gauge equivalent. This can only happen if the holonomies around any closed curve on $M_2$ are the same for all solutions. This, in turn, implies that $j \in \mZ$ and we can fix gauge transformations on the boundary in such a way that we only get contributions from the solution with $n=0$.  
 
Consequently, there is a unique $SO(3)$ gauge field solution for which the action is given by $I_{EM}^{SO(3), N} = \frac{G_N\beta  j^2}{r_0^3}$. For sufficiently large $j \gg 1$, this agrees with terms in the exponent in the sum over $j$ \eqref{eq:rep-to-SO3=instanton}. Since the $r$-dependence of the gauge field in \eqref{eq:SO3-gauge-field-sol-fixedF} is the same as that in \eqref{eq:SO3-instanton-solutions} we can once again check that when $j$ is sufficiently small that it does not backreact on $f(r)$,\footnote{$j(j+1) \ll (r_h/\ell_{Pl})^{4}$ as discussed in section \ref{sec:two-dimensional-gauge-fields}.} then Einstein's equations are indeed satisfied for the $4d$ metric ansatz when using the solution \eqref{eq:SO3-gauge-field-sol-fixedF}. 

\subsection{Uplift of the $SO(3)$ solution}
\label{sec:SO3-classical-solution}
      
In this section we will take the solution for the $SO(3)$ gauge field and show that it can be understood as a solution of the higher dimensional metric for small angular momentum. The KN solution is the unique solution with fixed $U(1)$ charge and angular momentum that also has a $U(1)$ spatial isometry \cite{Israel:1967wq, Carter:1971zc}. Therefore, by finding the angular momentum for the solutions analyzed in \ref{sec:SO3-classical-solution} in which we either fix the $SO(3)$ holonomy or the $SO(3)$ field strength we will determine the diffeomorphic equivalent KN solution.

 We saw above a solution for the gauge fields appears in the metric as 
\beq
 \delta g_{\mu \nu}^{SO(3)} dx^\mu dx^\nu = 2i r^2 \sin^2( \theta) \left( \a_1 + \frac{\a_2}{r^3}\right)d\tau d\phi
\eeq
to linear order in the angular momentum (i.e. no backreaction to $f(r)$), with respect to the charged black hole solution. The equation of motion for the four dimensional Einstein Maxwell theory is $G_{AB}\equiv R_{AB} -{\small \frac{1}{2}} g_{AB} R - {\small \frac{3}{L^2}} g_{AB} = 8 \pi G_N T_{AB}$ where $T_{AB} =\frac{1}{4 e^2}( F_{AC}F^C_B - g_{AB} F^2)$ is the stress tensor of the $U(1)$ gauge field. Expanding this to linear order in $\a_1$ and $\a_2$ we can check this corrections satisfies the equation of motion to linear order 
\beq
\frac{1}{8\pi G}\delta G_{\tau \phi} = \delta T_{\tau \phi} = \frac{Q^2}{32 \pi^2 r^2 } \left( \a_1 + \frac{\a_2}{r^3}\right) \sin^2 \theta,
\eeq
and all other components for both $\delta G$ and $\delta T$ vanish. The uniqueness of KN solution suggests \eqref{metext-KN-BHs-AdS} is the correct non-linear completion of this correction, written in a different gauge.

\bibliographystyle{utphys2}
{\small \bibliography{Biblio}{}}

\providecommand{\href}[2]{#2}\begingroup\raggedright\begin{thebibliography}{10}

\bibitem{Kunduri:2007vf}
H.~K. Kunduri, J.~Lucietti, and H.~S. Reall, ``{Near-horizon symmetries of
  extremal black holes},''
  \href{http://dx.doi.org/10.1088/0264-9381/24/16/012}{{\em Class. Quant.
  Grav.} {\bfseries 24} (2007) 4169--4190},
\href{http://arxiv.org/abs/0705.4214}{{\ttfamily arXiv:0705.4214 [hep-th]}}.

\bibitem{Preskill:1991tb}
J.~Preskill, P.~Schwarz, A.~D. Shapere, S.~Trivedi, and F.~Wilczek,
  ``{Limitations on the statistical description of black holes},''
\href{http://dx.doi.org/10.1142/S0217732391002773}{{\em Mod. Phys. Lett.}
  {\bfseries A6} (1991) 2353--2362}.

\bibitem{Maldacena:1998uz}
J.~M. Maldacena, J.~Michelson, and A.~Strominger, ``{Anti-de Sitter
  fragmentation},'' \href{http://dx.doi.org/10.1088/1126-6708/1999/02/011}{{\em
  JHEP} {\bfseries 02} (1999) 011},
\href{http://arxiv.org/abs/hep-th/9812073}{{\ttfamily arXiv:hep-th/9812073
  [hep-th]}}.

\bibitem{Page:2000dk}
D.~N. Page, ``{Thermodynamics of near extreme black holes},''
\href{http://arxiv.org/abs/hep-th/0012020}{{\ttfamily arXiv:hep-th/0012020
  [hep-th]}}.

\bibitem{Callan:1996dv}
C.~G. Callan and J.~M. Maldacena, ``{D-brane approach to black hole quantum
  mechanics},'' \href{http://dx.doi.org/10.1016/0550-3213(96)00225-8}{{\em
  Nucl. Phys.} {\bfseries B472} (1996) 591--610},
\href{http://arxiv.org/abs/hep-th/9602043}{{\ttfamily arXiv:hep-th/9602043
  [hep-th]}}.

\bibitem{Maldacena:1996ds}
J.~M. Maldacena and L.~Susskind, ``{D-branes and fat black holes},''
  \href{http://dx.doi.org/10.1016/0550-3213(96)00323-9}{{\em Nucl. Phys.}
  {\bfseries B475} (1996) 679--690},
\href{http://arxiv.org/abs/hep-th/9604042}{{\ttfamily arXiv:hep-th/9604042
  [hep-th]}}.

\bibitem{Maldacena:1997ih}
J.~M. Maldacena and A.~Strominger, ``{Universal low-energy dynamics for
  rotating black holes},''
  \href{http://dx.doi.org/10.1103/PhysRevD.56.4975}{{\em Phys. Rev.} {\bfseries
  D56} (1997) 4975--4983},
\href{http://arxiv.org/abs/hep-th/9702015}{{\ttfamily arXiv:hep-th/9702015
  [hep-th]}}.

\bibitem{HITZ}
M.~Heydeman, L.~V. Iliesiu, G.~J. Turiaci, and W.~Zhao, ``{The statistical
  mechanics of near-BPS black holes},''
  \href{http://arxiv.org/abs/2011.01953}{{\ttfamily arXiv:2011.01953
  [hep-th]}}.

\bibitem{Banerjee:2010qc}
S.~Banerjee, R.~K. Gupta, and A.~Sen, ``{Logarithmic Corrections to Extremal
  Black Hole Entropy from Quantum Entropy Function},''
  \href{http://dx.doi.org/10.1007/JHEP03(2011)147}{{\em JHEP} {\bfseries 03}
  (2011) 147},
\href{http://arxiv.org/abs/1005.3044}{{\ttfamily arXiv:1005.3044 [hep-th]}}.

\bibitem{Banerjee:2011jp}
S.~Banerjee, R.~K. Gupta, I.~Mandal, and A.~Sen, ``{Logarithmic Corrections to
  N=4 and N=8 Black Hole Entropy: A One Loop Test of Quantum Gravity},''
  \href{http://dx.doi.org/10.1007/JHEP11(2011)143}{{\em JHEP} {\bfseries 11}
  (2011) 143},
\href{http://arxiv.org/abs/1106.0080}{{\ttfamily arXiv:1106.0080 [hep-th]}}.

\bibitem{Sen:2011ba}
A.~Sen, ``{Logarithmic Corrections to N=2 Black Hole Entropy: An Infrared
  Window into the Microstates},''
  \href{http://dx.doi.org/10.1007/s10714-012-1336-5}{{\em Gen. Rel. Grav.}
  {\bfseries 44} no.~5, (2012) 1207--1266},
\href{http://arxiv.org/abs/1108.3842}{{\ttfamily arXiv:1108.3842 [hep-th]}}.

\bibitem{Sen:2012cj}
A.~Sen, ``{Logarithmic Corrections to Rotating Extremal Black Hole Entropy in
  Four and Five Dimensions},''
  \href{http://dx.doi.org/10.1007/s10714-012-1373-0}{{\em Gen. Rel. Grav.}
  {\bfseries 44} (2012) 1947--1991},
\href{http://arxiv.org/abs/1109.3706}{{\ttfamily arXiv:1109.3706 [hep-th]}}.

\bibitem{Loran:2010bd}
F.~Loran, M.~M. Sheikh-Jabbari, and M.~Vincon, ``{Beyond Logarithmic
  Corrections to Cardy Formula},''
  \href{http://dx.doi.org/10.1007/JHEP01(2011)110}{{\em JHEP} {\bfseries 01}
  (2011) 110}, \href{http://arxiv.org/abs/1010.3561}{{\ttfamily arXiv:1010.3561
  [hep-th]}}.

\bibitem{Hawking:1994ii}
S.~W. Hawking, G.~T. Horowitz, and S.~F. Ross, ``{Entropy, Area, and black hole
  pairs},'' \href{http://dx.doi.org/10.1103/PhysRevD.51.4302}{{\em Phys. Rev.}
  {\bfseries D51} (1995) 4302--4314},
\href{http://arxiv.org/abs/gr-qc/9409013}{{\ttfamily arXiv:gr-qc/9409013
  [gr-qc]}}.

\bibitem{Ghosh:2019rcj}
A.~Ghosh, H.~Maxfield, and G.~J. Turiaci, ``{A universal Schwarzian sector in
  two-dimensional conformal field theories},''
\href{http://arxiv.org/abs/1912.07654}{{\ttfamily arXiv:1912.07654 [hep-th]}}.

\bibitem{Teitelboim:1983ux}
C.~Teitelboim, ``{Gravitation and Hamiltonian Structure in Two Space-Time
  Dimensions},''
\href{http://dx.doi.org/10.1016/0370-2693(83)90012-6}{{\em Phys. Lett.}
  {\bfseries 126B} (1983) 41--45}.

\bibitem{Jackiw:1984je}
R.~Jackiw, ``{Lower Dimensional Gravity},''
\href{http://dx.doi.org/10.1016/0550-3213(85)90448-1}{{\em Nucl. Phys.}
  {\bfseries B252} (1985) 343--356}.

\bibitem{Iliesiu:2019lfc}
L.~V. Iliesiu, ``{On 2D gauge theories in Jackiw-Teitelboim gravity},''
\href{http://arxiv.org/abs/1909.05253}{{\ttfamily arXiv:1909.05253 [hep-th]}}.

\bibitem{Kapec:2019ecr}
D.~Kapec, R.~Mahajan, and D.~Stanford, ``{Matrix ensembles with global
  symmetries and 't Hooft anomalies from 2d gauge theory},''
\href{http://arxiv.org/abs/1912.12285}{{\ttfamily arXiv:1912.12285 [hep-th]}}.

\bibitem{Almheiri:2014cka}
A.~Almheiri and J.~Polchinski, ``{Models of AdS$_{2}$ backreaction and
  holography},'' \href{http://dx.doi.org/10.1007/JHEP11(2015)014}{{\em JHEP}
  {\bfseries 11} (2015) 014},
\href{http://arxiv.org/abs/1402.6334}{{\ttfamily arXiv:1402.6334 [hep-th]}}.

\bibitem{kitaevTalks}
A.~Kitaev, ``{Talks given at the Fundamental Physics Prize Symposium and KITP
  seminars},''.

\bibitem{Maldacena:2016hyu}
J.~Maldacena and D.~Stanford, ``{Remarks on the Sachdev-Ye-Kitaev model},''
  \href{http://dx.doi.org/10.1103/PhysRevD.94.106002}{{\em Phys. Rev.}
  {\bfseries D94} no.~10, (2016) 106002},
\href{http://arxiv.org/abs/1604.07818}{{\ttfamily arXiv:1604.07818 [hep-th]}}.

\bibitem{Jensen:2016pah}
K.~Jensen, ``{Chaos in AdS$_2$ Holography},''
  \href{http://dx.doi.org/10.1103/PhysRevLett.117.111601}{{\em Phys. Rev.
  Lett.} {\bfseries 117} no.~11, (2016) 111601},
\href{http://arxiv.org/abs/1605.06098}{{\ttfamily arXiv:1605.06098 [hep-th]}}.

\bibitem{Maldacena:2016upp}
J.~Maldacena, D.~Stanford, and Z.~Yang, ``{Conformal symmetry and its breaking
  in two dimensional Nearly Anti-de-Sitter space},''
  \href{http://dx.doi.org/10.1093/ptep/ptw124}{{\em PTEP} {\bfseries 2016}
  no.~12, (2016) 12C104},
\href{http://arxiv.org/abs/1606.01857}{{\ttfamily arXiv:1606.01857 [hep-th]}}.

\bibitem{Engelsoy:2016xyb}
J.~Engelsöy, T.~G. Mertens, and H.~Verlinde, ``{An investigation of AdS$_{2}$
  backreaction and holography},''
  \href{http://dx.doi.org/10.1007/JHEP07(2016)139}{{\em JHEP} {\bfseries 07}
  (2016) 139},
\href{http://arxiv.org/abs/1606.03438}{{\ttfamily arXiv:1606.03438 [hep-th]}}.

\bibitem{Jevicki:2016ito}
A.~Jevicki and K.~Suzuki, ``{Bi-Local Holography in the SYK Model:
  Perturbations},'' \href{http://dx.doi.org/10.1007/JHEP11(2016)046}{{\em JHEP}
  {\bfseries 11} (2016) 046},
\href{http://arxiv.org/abs/1608.07567}{{\ttfamily arXiv:1608.07567 [hep-th]}}.

\bibitem{Bagrets:2016cdf}
D.~Bagrets, A.~Altland, and A.~Kamenev, ``{Sachdev–Ye–Kitaev model as
  Liouville quantum mechanics},''
  \href{http://dx.doi.org/10.1016/j.nuclphysb.2016.08.002}{{\em Nucl. Phys.}
  {\bfseries B911} (2016) 191--205},
\href{http://arxiv.org/abs/1607.00694}{{\ttfamily arXiv:1607.00694
  [cond-mat.str-el]}}.

\bibitem{Stanford:2017thb}
D.~Stanford and E.~Witten, ``{Fermionic Localization of the Schwarzian
  Theory},'' \href{http://dx.doi.org/10.1007/JHEP10(2017)008}{{\em JHEP}
  {\bfseries 10} (2017) 008},
\href{http://arxiv.org/abs/1703.04612}{{\ttfamily arXiv:1703.04612 [hep-th]}}.

\bibitem{Mertens:2017mtv}
T.~G. Mertens, G.~J. Turiaci, and H.~L. Verlinde, ``{Solving the Schwarzian via
  the Conformal Bootstrap},''
  \href{http://dx.doi.org/10.1007/JHEP08(2017)136}{{\em JHEP} {\bfseries 08}
  (2017) 136},
\href{http://arxiv.org/abs/1705.08408}{{\ttfamily arXiv:1705.08408 [hep-th]}}.

\bibitem{Lam:2018pvp}
H.~T. Lam, T.~G. Mertens, G.~J. Turiaci, and H.~Verlinde, ``{Shockwave S-matrix
  from Schwarzian Quantum Mechanics},''
  \href{http://dx.doi.org/10.1007/JHEP11(2018)182}{{\em JHEP} {\bfseries 11}
  (2018) 182},
\href{http://arxiv.org/abs/1804.09834}{{\ttfamily arXiv:1804.09834 [hep-th]}}.

\bibitem{Goel:2018ubv}
A.~Goel, H.~T. Lam, G.~J. Turiaci, and H.~Verlinde, ``{Expanding the Black Hole
  Interior: Partially Entangled Thermal States in SYK},''
  \href{http://dx.doi.org/10.1007/JHEP02(2019)156}{{\em JHEP} {\bfseries 02}
  (2019) 156},
\href{http://arxiv.org/abs/1807.03916}{{\ttfamily arXiv:1807.03916 [hep-th]}}.

\bibitem{Kitaev:2018wpr}
A.~Kitaev and S.~J. Suh, ``{Statistical mechanics of a two-dimensional black
  hole},''
\href{http://arxiv.org/abs/1808.07032}{{\ttfamily arXiv:1808.07032 [hep-th]}}.

\bibitem{Yang:2018gdb}
Z.~Yang, ``{The Quantum Gravity Dynamics of Near Extremal Black Holes},''
  \href{http://dx.doi.org/10.1007/JHEP05(2019)205}{{\em JHEP} {\bfseries 05}
  (2019) 205},
\href{http://arxiv.org/abs/1809.08647}{{\ttfamily arXiv:1809.08647 [hep-th]}}.

\bibitem{Saad:2019lba}
P.~Saad, S.~H. Shenker, and D.~Stanford, ``{JT gravity as a matrix integral},''
\href{http://arxiv.org/abs/1903.11115}{{\ttfamily arXiv:1903.11115 [hep-th]}}.

\bibitem{Mertens:2019tcm}
T.~G. Mertens and G.~J. Turiaci, ``{Defects in Jackiw-Teitelboim Quantum
  Gravity},'' \href{http://dx.doi.org/10.1007/JHEP08(2019)127}{{\em JHEP}
  {\bfseries 08} (2019) 127},
\href{http://arxiv.org/abs/1904.05228}{{\ttfamily arXiv:1904.05228 [hep-th]}}.

\bibitem{Iliesiu:2019xuh}
L.~V. Iliesiu, S.~S. Pufu, H.~Verlinde, and Y.~Wang, ``{An exact quantization
  of Jackiw-Teitelboim gravity},''
\href{http://arxiv.org/abs/1905.02726}{{\ttfamily arXiv:1905.02726 [hep-th]}}.

\bibitem{Almheiri:2016fws}
A.~Almheiri and B.~Kang, ``{Conformal Symmetry Breaking and Thermodynamics of
  Near-Extremal Black Holes},''
  \href{http://dx.doi.org/10.1007/JHEP10(2016)052}{{\em JHEP} {\bfseries 10}
  (2016) 052},
\href{http://arxiv.org/abs/1606.04108}{{\ttfamily arXiv:1606.04108 [hep-th]}}.

\bibitem{Anninos:2017cnw}
D.~Anninos, T.~Anous, and R.~T. D'Agnolo, ``{Marginal deformations \& rotating
  horizons},'' \href{http://dx.doi.org/10.1007/JHEP12(2017)095}{{\em JHEP}
  {\bfseries 12} (2017) 095},
\href{http://arxiv.org/abs/1707.03380}{{\ttfamily arXiv:1707.03380 [hep-th]}}.

\bibitem{Sarosi:2017ykf}
G.~Sárosi, ``{AdS$_{2}$ holography and the SYK model},''
  \href{http://dx.doi.org/10.22323/1.323.0001}{{\em PoS} {\bfseries Modave2017}
  (2018) 001},
\href{http://arxiv.org/abs/1711.08482}{{\ttfamily arXiv:1711.08482 [hep-th]}}.

\bibitem{Nayak:2018qej}
P.~Nayak, A.~Shukla, R.~M. Soni, S.~P. Trivedi, and V.~Vishal, ``{On the
  Dynamics of Near-Extremal Black Holes},''
  \href{http://dx.doi.org/10.1007/JHEP09(2018)048}{{\em JHEP} {\bfseries 09}
  (2018) 048},
\href{http://arxiv.org/abs/1802.09547}{{\ttfamily arXiv:1802.09547 [hep-th]}}.

\bibitem{Moitra:2018jqs}
U.~Moitra, S.~P. Trivedi, and V.~Vishal, ``{Extremal and near-extremal black
  holes and near-CFT$_{1}$},''
  \href{http://dx.doi.org/10.1007/JHEP07(2019)055}{{\em JHEP} {\bfseries 07}
  (2019) 055},
\href{http://arxiv.org/abs/1808.08239}{{\ttfamily arXiv:1808.08239 [hep-th]}}.

\bibitem{Hadar:2018izi}
S.~Hadar, ``{Near-extremal black holes at late times, backreacted},''
  \href{http://dx.doi.org/10.1007/JHEP01(2019)214}{{\em JHEP} {\bfseries 01}
  (2019) 214},
\href{http://arxiv.org/abs/1811.01022}{{\ttfamily arXiv:1811.01022 [hep-th]}}.

\bibitem{Castro:2018ffi}
A.~Castro, F.~Larsen, and I.~Papadimitriou, ``{5D rotating black holes and the
  nAdS$_{2}$/nCFT$_{1}$ correspondence},''
  \href{http://dx.doi.org/10.1007/JHEP10(2018)042}{{\em JHEP} {\bfseries 10}
  (2018) 042},
\href{http://arxiv.org/abs/1807.06988}{{\ttfamily arXiv:1807.06988 [hep-th]}}.

\bibitem{Larsen:2018cts}
F.~Larsen and Y.~Zeng, ``{Black hole spectroscopy and AdS$_{2}$ holography},''
  \href{http://dx.doi.org/10.1007/JHEP04(2019)164}{{\em JHEP} {\bfseries 04}
  (2019) 164},
\href{http://arxiv.org/abs/1811.01288}{{\ttfamily arXiv:1811.01288 [hep-th]}}.

\bibitem{Moitra:2019bub}
U.~Moitra, S.~K. Sake, S.~P. Trivedi, and V.~Vishal, ``{Jackiw-Teitelboim
  Gravity and Rotating Black Holes},''
\href{http://arxiv.org/abs/1905.10378}{{\ttfamily arXiv:1905.10378 [hep-th]}}.

\bibitem{Sachdev:2019bjn}
S.~Sachdev, ``{Universal low temperature theory of charged black holes with
  AdS$_2$ horizons},'' \href{http://dx.doi.org/10.1063/1.5092726}{{\em J. Math.
  Phys.} {\bfseries 60} no.~5, (2019) 052303},
\href{http://arxiv.org/abs/1902.04078}{{\ttfamily arXiv:1902.04078 [hep-th]}}.

\bibitem{Hong:2019tsx}
J.~Hong, F.~Larsen, and J.~T. Liu, ``{The scales of black holes with nAdS$_{2}$
  geometry},'' \href{http://dx.doi.org/10.1007/JHEP10(2019)260}{{\em JHEP}
  {\bfseries 10} (2019) 260},
\href{http://arxiv.org/abs/1907.08862}{{\ttfamily arXiv:1907.08862 [hep-th]}}.

\bibitem{Castro:2019crn}
A.~Castro and V.~Godet, ``{Breaking away from the near horizon of extreme
  Kerr},''
\href{http://arxiv.org/abs/1906.09083}{{\ttfamily arXiv:1906.09083 [hep-th]}}.

\bibitem{Charles:2019tiu}
A.~M. Charles and F.~Larsen, ``{A One-Loop Test of the
  near-AdS$_2$/near-CFT$_1$ Correspondence},''
\href{http://arxiv.org/abs/1908.03575}{{\ttfamily arXiv:1908.03575 [hep-th]}}.

\bibitem{Braden:1990hw}
H.~W. Braden, J.~D. Brown, B.~F. Whiting, and J.~W. York, Jr., ``{Charged black
  hole in a grand canonical ensemble},''
\href{http://dx.doi.org/10.1103/PhysRevD.42.3376}{{\em Phys. Rev.} {\bfseries
  D42} (1990) 3376--3385}.

\bibitem{Hawking:1995ap}
S.~W. Hawking and S.~F. Ross, ``{Duality between electric and magnetic black
  holes},'' \href{http://dx.doi.org/10.1103/PhysRevD.52.5865}{{\em Phys. Rev.}
  {\bfseries D52} (1995) 5865--5876},
\href{http://arxiv.org/abs/hep-th/9504019}{{\ttfamily arXiv:hep-th/9504019
  [hep-th]}}.

\bibitem{Henningson:1998ey}
M.~Henningson and K.~Skenderis, ``{Holography and the Weyl anomaly},''
  \href{http://dx.doi.org/10.1002/(SICI)1521-3978(20001)48:1/3<125::AID-PROP125>3.0.CO;2-B,
  10.1002/(SICI)1521-3978(20001)48:1/3<125::AID-PROP125>3.3.CO;2-2}{{\em
  Fortsch. Phys.} {\bfseries 48} (2000) 125--128},
\href{http://arxiv.org/abs/hep-th/9812032}{{\ttfamily arXiv:hep-th/9812032
  [hep-th]}}.

\bibitem{Balasubramanian:1999re}
V.~Balasubramanian and P.~Kraus, ``{A Stress tensor for Anti-de Sitter
  gravity},'' \href{http://dx.doi.org/10.1007/s002200050764}{{\em Commun. Math.
  Phys.} {\bfseries 208} (1999) 413--428},
\href{http://arxiv.org/abs/hep-th/9902121}{{\ttfamily arXiv:hep-th/9902121
  [hep-th]}}.

\bibitem{Chamblin:1999tk}
A.~Chamblin, R.~Emparan, C.~V. Johnson, and R.~C. Myers, ``{Charged AdS black
  holes and catastrophic holography},''
  \href{http://dx.doi.org/10.1103/PhysRevD.60.064018}{{\em Phys. Rev.}
  {\bfseries D60} (1999) 064018},
\href{http://arxiv.org/abs/hep-th/9902170}{{\ttfamily arXiv:hep-th/9902170
  [hep-th]}}.

\bibitem{Jafferis:2017zna}
D.~Jafferis, B.~Mukhametzhanov, and A.~Zhiboedov, ``{Conformal Bootstrap At
  Large Charge},'' \href{http://dx.doi.org/10.1007/JHEP05(2018)043}{{\em JHEP}
  {\bfseries 05} (2018) 043},
\href{http://arxiv.org/abs/1710.11161}{{\ttfamily arXiv:1710.11161 [hep-th]}}.

\bibitem{Salam:1984cj}
A.~Salam and E.~Sezgin, ``{Chiral Compactification on Minkowski$\times S^2$ of
  $N=2$ Einstein-Maxwell Supergravity in Six-Dimensions},''
  \href{http://dx.doi.org/10.1016/0370-2693(84)90589-6}{{\em Phys. Lett.}
  {\bfseries 147B} (1984) 47}.
[,47(1984)].

\bibitem{Michelson:1999kn}
J.~Michelson and M.~Spradlin, ``{Supergravity spectrum on AdS(2)$\times
  S^2$},'' \href{http://dx.doi.org/10.1088/1126-6708/1999/09/029}{{\em JHEP}
  {\bfseries 09} (1999) 029},
\href{http://arxiv.org/abs/hep-th/9906056}{{\ttfamily arXiv:hep-th/9906056
  [hep-th]}}.

\bibitem{GibbonsPope}
G.~W. Gibbons and C.~N. Pope, ``{Consistent S$^2$ Pauli reduction of
  six-dimensional chiral gauged Einstein-Maxwell supergravity},''
  \href{http://dx.doi.org/10.1016/j.nuclphysb.2004.07.016}{{\em Nucl. Phys.}
  {\bfseries B697} (2004) 225--242},
\href{http://arxiv.org/abs/hep-th/0307052}{{\ttfamily arXiv:hep-th/0307052
  [hep-th]}}.

\bibitem{Maxfield:2020ale}
H.~Maxfield and G.~J. Turiaci, ``{The path integral of 3D gravity near
  extremality; or, JT gravity with defects as a matrix integral},''
  \href{http://arxiv.org/abs/2006.11317}{{\ttfamily arXiv:2006.11317
  [hep-th]}}.

\bibitem{Grumiller:2007ju}
D.~Grumiller and R.~McNees, ``{Thermodynamics of black holes in two (and
  higher) dimensions},''
  \href{http://dx.doi.org/10.1088/1126-6708/2007/04/074}{{\em JHEP} {\bfseries
  04} (2007) 074},
\href{http://arxiv.org/abs/hep-th/0703230}{{\ttfamily arXiv:hep-th/0703230
  [HEP-TH]}}.

\bibitem{Emparan:1999pm}
R.~Emparan, C.~V. Johnson, and R.~C. Myers, ``{Surface terms as counterterms in
  the AdS / CFT correspondence},''
  \href{http://dx.doi.org/10.1103/PhysRevD.60.104001}{{\em Phys. Rev.}
  {\bfseries D60} (1999) 104001},
\href{http://arxiv.org/abs/hep-th/9903238}{{\ttfamily arXiv:hep-th/9903238
  [hep-th]}}.

\bibitem{Kitaev:2017awl}
A.~Kitaev and S.~J. Suh, ``{The soft mode in the Sachdev-Ye-Kitaev model and
  its gravity dual},'' \href{http://dx.doi.org/10.1007/JHEP05(2018)183}{{\em
  JHEP} {\bfseries 05} (2018) 183},
\href{http://arxiv.org/abs/1711.08467}{{\ttfamily arXiv:1711.08467 [hep-th]}}.

\bibitem{Sen:2005wa}
A.~Sen, ``{Black hole entropy function and the attractor mechanism in higher
  derivative gravity},''
  \href{http://dx.doi.org/10.1088/1126-6708/2005/09/038}{{\em JHEP} {\bfseries
  09} (2005) 038},
\href{http://arxiv.org/abs/hep-th/0506177}{{\ttfamily arXiv:hep-th/0506177
  [hep-th]}}.

\bibitem{Larsen:2014bqa}
F.~Larsen and P.~Lisbao, ``{Quantum Corrections to Supergravity on AdS$_2\times
  S^2$},'' \href{http://dx.doi.org/10.1103/PhysRevD.91.084056}{{\em Phys. Rev.}
  {\bfseries D91} no.~8, (2015) 084056},
\href{http://arxiv.org/abs/1411.7423}{{\ttfamily arXiv:1411.7423 [hep-th]}}.

\bibitem{gel1960integration}
I.~M. Gel'fand and A.~M. Yaglom, ``Integration in functional spaces and its
  applications in quantum physics,'' {\em Journal of Mathematical Physics}
  {\bfseries 1} no.~1, (1960) 48--69.

\bibitem{Maldacena:2019cbz}
J.~Maldacena, G.~J. Turiaci, and Z.~Yang, ``{Two dimensional Nearly de Sitter
  gravity},''
\href{http://arxiv.org/abs/1904.01911}{{\ttfamily arXiv:1904.01911 [hep-th]}}.

\bibitem{Gubser:2008px}
S.~S. Gubser, ``{Breaking an Abelian gauge symmetry near a black hole
  horizon},'' \href{http://dx.doi.org/10.1103/PhysRevD.78.065034}{{\em Phys.
  Rev.} {\bfseries D78} (2008) 065034},
\href{http://arxiv.org/abs/0801.2977}{{\ttfamily arXiv:0801.2977 [hep-th]}}.

\bibitem{Gubser:2008wv}
S.~S. Gubser and S.~S. Pufu, ``{The Gravity dual of a p-wave superconductor},''
  \href{http://dx.doi.org/10.1088/1126-6708/2008/11/033}{{\em JHEP} {\bfseries
  11} (2008) 033},
\href{http://arxiv.org/abs/0805.2960}{{\ttfamily arXiv:0805.2960 [hep-th]}}.

\bibitem{Liu:2009dm}
H.~Liu, J.~McGreevy, and D.~Vegh, ``{Non-Fermi liquids from holography},''
  \href{http://dx.doi.org/10.1103/PhysRevD.83.065029}{{\em Phys. Rev.}
  {\bfseries D83} (2011) 065029},
\href{http://arxiv.org/abs/0903.2477}{{\ttfamily arXiv:0903.2477 [hep-th]}}.

\bibitem{Faulkner:2009wj}
T.~Faulkner, H.~Liu, J.~McGreevy, and D.~Vegh, ``{Emergent quantum criticality,
  Fermi surfaces, and AdS(2)},''
  \href{http://dx.doi.org/10.1103/PhysRevD.83.125002}{{\em Phys. Rev.}
  {\bfseries D83} (2011) 125002},
\href{http://arxiv.org/abs/0907.2694}{{\ttfamily arXiv:0907.2694 [hep-th]}}.

\bibitem{Berkooz:2016cvq}
M.~Berkooz, P.~Narayan, M.~Rozali, and J.~Simon, ``{Higher Dimensional
  Generalizations of the SYK Model},''
  \href{http://dx.doi.org/10.1007/JHEP01(2017)138}{{\em JHEP} {\bfseries 01}
  (2017) 138},
\href{http://arxiv.org/abs/1610.02422}{{\ttfamily arXiv:1610.02422 [hep-th]}}.

\bibitem{Turiaci:2017zwd}
G.~Turiaci and H.~Verlinde, ``{Towards a 2d QFT Analog of the SYK Model},''
  \href{http://dx.doi.org/10.1007/JHEP10(2017)167}{{\em JHEP} {\bfseries 10}
  (2017) 167},
\href{http://arxiv.org/abs/1701.00528}{{\ttfamily arXiv:1701.00528 [hep-th]}}.

\bibitem{Murugan:2017eto}
J.~Murugan, D.~Stanford, and E.~Witten, ``{More on Supersymmetric and 2d
  Analogs of the SYK Model},''
  \href{http://dx.doi.org/10.1007/JHEP08(2017)146}{{\em JHEP} {\bfseries 08}
  (2017) 146},
\href{http://arxiv.org/abs/1706.05362}{{\ttfamily arXiv:1706.05362 [hep-th]}}.

\bibitem{Almheiri:2019qdq}
A.~Almheiri, T.~Hartman, J.~Maldacena, E.~Shaghoulian, and A.~Tajdini,
  ``{Replica Wormholes and the Entropy of Hawking Radiation},''
\href{http://arxiv.org/abs/1911.12333}{{\ttfamily arXiv:1911.12333 [hep-th]}}.

\bibitem{Penington:2019kki}
G.~Penington, S.~H. Shenker, D.~Stanford, and Z.~Yang, ``{Replica wormholes and
  the black hole interior},''
\href{http://arxiv.org/abs/1911.11977}{{\ttfamily arXiv:1911.11977 [hep-th]}}.

\bibitem{Israel:1967wq}
W.~Israel, ``{Event horizons in static vacuum space-times},''
\href{http://dx.doi.org/10.1103/PhysRev.164.1776}{{\em Phys. Rev.} {\bfseries
  164} (1967) 1776--1779}.

\bibitem{Carter:1971zc}
B.~Carter, ``{Axisymmetric Black Hole Has Only Two Degrees of Freedom},''
\href{http://dx.doi.org/10.1103/PhysRevLett.26.331}{{\em Phys. Rev. Lett.}
  {\bfseries 26} (1971) 331--333}.

\end{thebibliography}\endgroup

\end{document}